\documentclass[number,letterpaper,1p]{elsarticle}

\usepackage{graphicx}

\usepackage{amssymb}
\usepackage{amsmath,amsthm}

\journal{Computational Physics}

\usepackage{lmodern}

\def\R{\mathbb{R}}

\theoremstyle{plain}

\newtheorem{lem}{Lemma}[section]
\newtheorem*{thm*}{Theorem}
\newtheorem*{lem*}{Lemma}

\newtheorem*{lem21}{Lemma~\ref{thm2.1}}
\newtheorem*{lem22}{Lemma~\ref{thm2.2}}
\newtheorem*{lem23}{Lemma~\ref{thm2.3}}

\theoremstyle{remark}

\begin{document}

\begin{frontmatter}



\title{Deterministic Solution of the Boltzmann Equation Using Discontinuous Galerkin Discretizations in Velocity Space}


\author[csun,wpafb]{A.~Alekseenko\corref{cor1}}
\ead{Alexander.Alekseenko@csun.edu}

\author[wpafb]{E.~Josyula}
\ead{Eswar.Josyula@wpafb.af.mil}

\address[csun]{Department of Mathematics, California State University Northridge, 
Nortridge, CA, USA}

\address[wpafb]{Air Force Research Laboratory, Wright-Patterson AFB, OH, USA}

\cortext[cor1]{Corresponding author}

\begin{abstract}
We present a new deterministic approach for the solution of the Boltzmann kinetic equation based on nodal 
discontinuous Galerkin (DG) discretizations in velocity space. In the new approach the collision operator 
has the form of a bilinear operator with pre-computed kernel; its evaluation requires $O(n^5)$ operations 
at every point of the phase space where $n$ is the number of degrees of freedom in one velocity dimension. 
The method is generalized to any molecular potential. Results of numerical simulations are presented 
for the problem of spatially homogeneous relaxation for the hard spheres potential. Comparison with the method of Direct 
Simulation Monte Carlo (DSMC) showed excellent agreement.
\end{abstract}


\end{frontmatter}



\section{Introduction}

Being central to gas dynamics the Boltzmann equation has the capacity to describe gas flows in 
regimes from continuum to rarefied. Its descriptive power is derived from the microscopic 
probabilistic representation of gas by a space and time dependent velocity distribution function 
of a large collection of particles. The particles interact according to known potentials 
producing a change in the distribution function that is modelled by the five fold (three velocity and two spatial integrals) 
Boltzmann collision integral. The Boltzmann equation is one of the most intensely studied subjects over the last  
fifty years. Analytic solutions to this equation have been constructed for simple geometries and in special 
cases of molecular potentials. However, the complexity of the equation suggests that solutions to 
applications arising in engineering and physics with complex boundaries and complex gas-to-gas and 
gas-to-surface interactions can only be computed approximately, using 
numerical techniques. The costs associated with the direct evaluation of the collision integral, however, are very 
high even with the most advanced discretization methods. As a result, the Boltzmann equation 
is rarely solved directly in multidimensional applications. Instead, alternative techniques such as direct simulation Monte 
Carlo (DSMC) methods, see e.g., \cite{Bird1994}, are used for simulating engineering 
applications. However statistical noise that is inherent to DSMC methods 
makes it cumbersome to couple these methods to deterministic models, for example, to the Navier-Stokes equations. 
Time dependent problems represent an additional challenge since the stochastic noise can propagate and perturb the 
continuum solution. Also, DSMC methods may become prohibitively expensive for the simulation of the
low speed flows where the flow velocity is less than the mean molecular
thermal velocity. It requires a very large statistical sample in this case to keep the 
statistical noise from overpowering the signal. To overcome these difficulties 
several simplified deterministic approaches were developed. In particular, 
Lattice-Boltzmann methods \cite{Mohamad2011}, the
discrete velocity methods \cite{Buet1996}, the method of model kinetic equations \cite{M00,Titarev2012}, the
method of moments and the extended hydrodynamics approach \cite{Struchtrup2005} are used to obtain approximate 
deterministic solutions to the Boltzmann equation. However, analytical error evaluation of these models is not 
straightforward. It is therefore 
important to develop methods capable to solve the full Boltzmann equation directly. Such methods can be used 
both for validation and for obtaining solutions where approximate techniques fail to be accurate. 

Historically, however, one can count only a handful of successful attempts at the solution of the full Boltzmann 
equation. In \cite{Tcher2003} Tcheremissine presented an approach for fast evaluation of the collision integral 
based on a uniform discrete ordinate velocity discretization. In this approach the collision integral is written in the form 
of an eight fold integral (six velocity and two spatial integrals) using Dirac delta-function formalism. Conservation 
of mass, momentum and energy is achieved by using a special interpolation of the velocity distribution function 
at the off-grid values of post-collision velocities. The efficient evaluation of the multidimensional integral 
is accomplished using quasi-stochastic Korobov integration. Tcheremissine's approach was successfully applied to 
simulations of the Boltzmann equation in more than one spatial dimensions \cite{KAAFZ2007,Aristov200449,josyula:017101} (however, see also 
\cite{AristovZabelok2002}). On the other hand, being not fully deterministic method, its accuracy is not easy to 
evaluate. In particular, it is anticipated that accuracy of Korobov schemes deteriorates on non-smooth solutions, 
such as ones arising in problems with gas-to-surface interactions. In \cite{BobRjas1999} Bobylev and Rjasanow developed a 
convolution formulation for the collision operator using Fourier transform. Their approach 
avoids interpolation of the solution in the off-grid points. Efficiency is achieved by using fast Fourier transform to 
evaluate integrals in velocity variable. However, Fourier decomposition of the distribution function is 
expected to require a large number of spectral degrees of freedom to accurately capture bimodal solutions that often appear 
in shock waves. Thus the methods based on the Fourier transform (as well as methods using other globally supported basis functions)
will not be easily adaptable for such problems. An approach based on the uniform piece-wise constant approximation 
of the distribution function was proposed by Aristov, see e.g., \cite{Aristov2001}. By 
exploring the simple form of the discrete distribution function and re-ordering quadrature rules the author transforms 
the discrete collision integral to the form of a bilinear operator. Moreover, the author establishes 
analytical expressions for components of the kernel of the collision integral in the case of hard spheres potential, 
e.g.,  \cite{AristovZabelok2002,Aristov2001}. Other examples of solutions of the Boltzmann equations include 
\cite{PareschiRusso2000,Varghese2007,PanferovHeintz2002}  and most 
recently \cite{MalkovIvanov2011,Morris20111265,MalkovIvanovPoleshkin2012,GhiroldiGibelli2012,AlekseenkoJosyula2012,MouhotParechiRey2012}. 

In \cite{Alekseenko2011,AGG12} it was shown that high order DG approximations in velocity space can be very accurate in 
preserving mass, momentum and energy in the discrete solution even if the solution is rough. In addition, DG methods 
are well suited for adaptive techniques and parallel implementation. This motivated the authors to develop a nodal DG discretization 
to the full Boltzmann equation. Our nodal DG basis in the velocity space is constructed from Lagrange basis functions on Gauss 
quadrature nodes \cite{HesthavenWarburtoin2007}. The resulting discretized form of the kinetic transport equation is equivalent 
to the discrete ordinate method. Using some well-known identities we rewrite the Galerkin projection of the Boltzmann 
collision operator in the form of a bilinear operator with a symmetric pre-computed kernel. To the knowledge of the authors, 
this form of the collision 
operator first appeared in the context of the method of moments, see, e.g., \cite{Bobylev1997}. Its importance for the direct 
discretization of the Boltzmann equation was first recognized by Pareschi and Perthame \cite{PareschiPerthame1996} 
(see also \cite{PareschiRusso2000}) who noticed that the resulting scheme requires $O(n^6)$ multiplications, where $n$ is the 
number of degrees of freedom in one velocity dimension. This is considerably less multiplications than in other spectral 
techniques (see the discussion in \cite{Pareschi2003,MouhotParechiRey2012}). While our form of the collision operator is equivalent 
to that of \cite{PareschiPerthame1996}, our approach requires only $O(n^5)$ multiplications. The savings come from the fact that 
we use locally supported basis functions and that the resulting pre-computed kernels are sparse. Preliminary results 
on this method appeared in \cite{AlekseenkoJosyula2012}. In this work we establish some useful symmetry properties 
of the collision kernel, perform additional study of the algorithm computational complexity and accuracy and present a 
comparison of the new deterministic method with the DSMC method \cite{Boyd1991411} for the problem of spatially homogeneous 
relaxation. 

The paper is organized as follows. Section~\ref{secN2} presents the kinetic framework and the Boltzmann equation. 
Section~\ref{secN3} describes the nodal DG discretization in the velocity space. The symmetric form of the Galerkin projection 
of the Boltzmann collision operator is introduced in Section~\ref{secN4}. The fully discrete form of the collision operator is discussed 
in Section~{\ref{secN5}}. The results of numerical simulations for the problem of spatially homogeneous relaxation
are given in Section~{\ref{secN6}}.

\section{The Boltzmann equation}
\label{secN2}

In the kinetic approach the gas is described using the molecular velocity distribution function
$f(t,\vec{x},\vec{v})$ which is defined by the following property: $f(t,\vec{x},\vec{v})d\vec{x}\,d\vec{v}$ 
gives the number of molecules that are contained in the box with the volume $d\vec{x}$ around point $\vec{x}$
whose velocities are contained in a box of volume $d\vec{v}$ around point $\vec{v}$.
Here by $d\vec{x}$ and $d\vec{v}$ we denote the volume elements $dx\, dy\, dz$ and $du\, dv\, dw$, correspondingly.
Evolution of the molecular distribution function is governed by the 
Boltzmann equation, which in the case of one component 
atomic gas has the form (see, for example \cite{Kogan1969,Cercignani2000})
\begin{equation}
\label{boltzm1}
\frac{\partial}{\partial t} f(t,\vec{x},\vec{v}) + \vec{v}\cdot \nabla_{x} f(t,\vec{x},\vec{v}) = I[f](t,\vec{x},\vec{v}).
\end{equation}
Here $I[f](t,\vec{x},\vec{v})$ is the molecular collision operator. In most instances, it is sufficient to only consider 
binary collisions between molecules. In this case the collision operator takes the 
form
\begin{equation}
\label{boltzm2}
I[f](t,\vec{x},\vec{v}) = \int_{\R^3}\int_{0}^{2\pi} \int_{0}^{b_{\ast}}
(f(t,\vec{x},\vec{v}')f(t,\vec{x},\vec{v}'_{1})-
f(t,\vec{x},\vec{v})f(t,\vec{x},\vec{v}_{1})) |\vec{g}|b\, db\, d\varphi\, d \vec{v}_{1},
\end{equation}
where $\vec{v}$ and $\vec{v}_{1}$ are the pre-collision and $\vec{v}'$ and $\vec{v}'_{1}$ are the 
post-collision velocities of a pair of particles, $\vec{g}=\vec{v}-\vec{v}_{1}$, $b$ is the distance 
of closest approach (the separation of the unperturbed trajectories) and $\varphi$ is the angle between the 
collision plane and some reference plane. Evaluation of the collision operator represents a considerable 
difficulty in the numerical solution of the Boltzmann equation. In the following sections we will develop
a high-order method for discretization of the Boltzmann equation in the velocity variable and design an 
algorithm for computing the collision operator based on this discretization.

\section{Discontinuous Galerkin velocity discretization}
\label{secN3}

Let us describe the DG velocity discretization that will be employed.
We select a rectangular parallelepiped in the velocity space that is 
sufficiently large so that contributions of the molecular distribution 
function to first few moments outside of this parallelepiped are negligible. 
In most cases, some a-priori knowledge about the problem is available and  
such parallelepiped can be selected. We partition this region into rectangular
parallelepipeds $K_{j}$. In this paper, only uniform partitions are considered; 
the advantages of using uniform partitions are explained in the next section. However, 
most of the approach carry over to non-uniform partitions and extensions to
hierarchical and overlapping meshes are straightforward. On each element $K_{j}$, 
$j=1,\ldots,M$ we introduce a finite dimensional functional basis 
$\phi(\vec{u})^{j}_{i}$, $i=1,\ldots s$. Notice that in general different 
approximation spaces can be used on different $K_{j}$. Thus the number of 
basis functions $s$ may be different for different velocity cells. However, 
the implementation of the method presented in this paper uses the same basis 
functions on all element to save on computational storage.

We define numbers $s_u$, $s_{v}$, and $s_{w}$ that determine the orders of the polynomial 
basis functions in components of velocity $u$, $v$, and $w$, respectively. Let
$K_{j}=[u^{j}_{L},u^{j}_{R}]\times[v^{j}_{L},v^{j}_{R}]\times[w^{j}_{L},w^{j}_{R}]$.
The basis functions are constructed as follows. We introduce nodes of the Gauss
quadratures of orders $s_u$, $s_{v}$, and $s_{w}$ on each of the intervals $[u^{j}_{L},u^{j}_{R}]$,
$[v^{j}_{L},v^{j}_{R}]$, and $[w^{j}_{L},w^{j}_{R}]$, respectively.  Let these nodes be 
denoted 
$\kappa^{j,u}_{i}$, $i=1,s_{u}$,
$\kappa^{j,v}_{i}$, $i=1,s_{v}$, and 
$\kappa^{j,w}_{i}$, $i=1,s_{w}$. 
We define one-dimensional Lagrange basis functions as follows, see e.g., \cite{HesthavenWarburtoin2007}, 
\begin{equation}
\label{eq01}
\phi^{j;u}_{i}(u)=\prod_{{l=1,s^{u}} \atop {l\neq i}} \frac{\kappa^{j,u}_{l}-u}{\kappa^{j,u}_{l}-\kappa^{j,u}_{i}}\, ,\quad 
\phi^{j;v}_{i}(v)=\prod_{{m=1,s^{v}} \atop {m\neq i}} \frac{\kappa^{j,v}_{m}-v}{\kappa^{j,v}_{m}-\kappa^{j,v}_{i}}\, ,\quad 
\phi^{j;w}_{i}(w)=\prod_{{n=1,s^{w}} \atop {m\neq i}} \frac{\kappa^{j,w}_{n}-u}{\kappa^{j,w}_{n}-\kappa^{j,w}_{i}}\, .
\end{equation}
The three-dimensional basis functions are defined as 
$\phi^{j}_{i}(\vec{v})=\phi^{j;u}_{l}(u)\phi^{j;v}_{m}(v)\phi^{j;w}_{n}(w)$, where
$i=1,\ldots,s=s_{u}s_{v}s_{w}$ is the index running through all 
combinations of $l$, $n$, and $m$. (In the implementation discussed in this paper, 
$i$ is computed using the following formula $i=(l-1)*s_{v}*s_{w}+(m-1)*s_{w}+n$.) 

\begin{lem} 
The following identities hold for basis functions $\phi^{j}_{i}(\vec{v})$:
\begin{equation}
\label{eq1.1} 
\int_{K_{j}} \phi^{j}_{p}(\vec{v})\phi^{j}_{q}(\vec{v})\, d\vec{v} = \frac{\Delta\vec{v}^{j}}{8}\omega_{p}\delta_{pq},
\qquad 
\int_{K_{j}} \vec{v}\phi^{j}_{p}(\vec{v})\phi^{j}_{q}(\vec{v})\, d\vec{v} 
= \frac{\Delta\vec{v}^{j}}{8}\vec{v}^{j}_{p}\omega_{p}\delta_{pq},
\end{equation}
where $\Delta\vec{v}^{j}= (u^{j}_{R}-u^{j}_{L})(v^{j}_{R}-v^{j}_{L}) (w^{j}_{R}-w^{j}_{L})$, 
$\omega_{p}:=\omega^{s_{u}}_{l} \omega^{s_{v}}_{m} \omega^{s_{w}}_{n}$, and 
$\omega^{s_{u}}_{l}$,  $\omega^{s_{v}}_{m}$, and $\omega^{s_{w}}_{n}$ are the weights
of the Gauss quadratures of orders $s_u$, $s_v$, and $s_w$, respectively 
and indices $l$, $n$, and $m$ of one dimensional basis functions correspond to 
the three-dimensional basis function
$\phi^{j}_{p}(\vec{v})=\phi^{j;u}_{l}(u)\phi^{j;v}_{m}(v)\phi^{j;w}_{n}(w)$, 
and the vector $\vec{v}^{j}_{p}=(\kappa^{j,u}_{l},\kappa^{j,v}_{m},\kappa^{j,w}_{n})$. 
\end{lem}

The lemma follows by re-writing the three-dimensional integral as 
an iterative integral and by reviewing the integrals of products of 
one-dimensional basis functions. The orthogonality of one-dimensional 
basis functions in each variable follows by replacing the 
one-dimensional integrals with Gauss quadratures on $s^{u}$ nodes 
(similarly, $s^{v}$ ans $s^w$ nodes) and recalling that these quadratures 
are precise on polynomials of degrees at most $2s^{u}-1$ (similarly, $2s^{v}-1$ and 
$2s^{w}-1$) and by recalling that the constructed basis functions 
vanish on all nodes but one at which they are equal to one. 

We assume that on each $K_{j}$ the solution to the Boltzmann equation 
is sought in the form 
\begin{equation}
\label{eq1.2}
f(t,\vec{x},\vec{v})|_{K_{j}} = \sum_{i=1,s} f_{i;j}(t,\vec{x})\phi^{j}_{i}(\vec{v})
\end{equation}
The discontinuous Galerkin (DG) velocity discretization that we shall study results by
substituting the representation (\ref{eq1.2}) into (\ref{boltzm1}) and multiplying 
the result by a test basis function and integrating over $K_{j}$. Repeating this 
for all $K_{j}$ and using the identities (\ref{eq1.1}) we arrive at
\begin{equation}
\label{discveloblzm}
\partial_{t} f_{i;j}(t,\vec{x}) + \vec{v}^{j}_{i}\cdot \nabla_{x} f_{i;j}(t,\vec{x}) =
I_{\phi^{j}_{i}}
\end{equation}
where $I_{\phi^{j}_{i}}$ is the projection of the collision operator 
on the basis function $\phi^{j}_{i}(\vec{v})$:
\begin{equation}
\label{projcoll}
I_{\phi^{j}_{i}} = \frac{8}{\omega_{i} \Delta \vec{v}^j} \int_{K_{j}}\phi^{j}_{i}(\vec{v}) I[f](t,\vec{x},\vec{v})\, d\vec{v}\, .
\end{equation} 
Notice that our velocity discretization is still incomplete because 
we have to specify how to evaluate the projection of the integral 
collision term. This is described in the next section. We however 
want to emphasize the simplicity of the obtained discrete velocity 
formulation. Indeed, the transport part of (\ref{discveloblzm}) has 
the complexity of a discrete ordinate formulation. A formulation with similar 
properties has been presented in Gobbert and Cale \cite{GobbertCale2007}. Their 
Galerkin formulation, however, uses global basis functions of high 
order Hermite's polynomials. Formulation (\ref{eq1.2}) therefore extends 
their approach.

\section{Reformulation of the Galerkin projection of the collision operator}
\label{secN4}

We will now introduce the formalism that will be used for evaluating 
the DG projection of the collision operator. We notice that $\phi^{j}_{i}(\vec{v})$ can 
be extended by zero to the entire $\R^3$. Then 
\begin{equation}
\label{eq2.1}
I_{\phi^{j}_{i}}=\frac{8}{\omega_{i} \Delta \vec{v}^j}\int_{R^3}\phi^{j}_{i}(\vec{v}) \int_{R^3}
\int_{0}^{2\pi} \int_{0}^{b_{\ast}}
(f(t,\vec{x},\vec{v}')f(t,\vec{x},\vec{v}_{1}')
-f(t,\vec{x},\vec{v})f(t,\vec{x},\vec{v}_{1})) 
|\vec{g}| b\, db\, d\varepsilon\, d\vec{v}_{1}\, d\vec{v}\, .
\end{equation}
Using symmetry properties of the collision operator 
(e.g., \cite{Kogan1969}, Section 2.4), 
we can replace the last expression with
\begin{equation}
\label{eq2.2}
I_{\phi^{j}_{i}}=\frac{8}{\omega_{i} \Delta \vec{v}^j}\int_{R^3}\int_{R^3}\frac{1}{2}
\int_{0}^{2\pi}\int_{0}^{b_{\ast}}
(\phi^{j}_{i}(\vec{v}') +\phi^{j}_{i}(\vec{v}'_{1})
-\phi^{j}_{i}(\vec{v})  -\phi^{j}_{i}(\vec{v}_{1}))
f(t,\vec{x},\vec{v})f(t,\vec{x},\vec{v}_{1}) |\vec{g}| b\, db\, d\varepsilon\, d\vec{v}_{1}\, d\vec{v}\, .
\end{equation}
The first principles of the kinetic theory imply that  
changes in $f(t,\vec{x},\vec{v})$ and $f(t,\vec{x},\vec{v}_{1})$ with 
respect to $\vec{x}$ are extremely small at distances of a few $b^{\ast}$, 
see e.g., \cite{Struchtrup2005}, Section~3.1.4. We will neglect these 
changes and therefore will assume that values of $\vec{x}$ in $f(t,\vec{x},\vec{v})$ 
and $f(t,\vec{x},\vec{v}_{1})$ are independent of the impact parameters $b$ and $\epsilon$. With this assumption, 
$f(t,\vec{x},\vec{v})$ and $f(t,\vec{x},\vec{v}_{1})$ can be removed from under the 
integrals in $b$ and $\epsilon$ in (\ref{eq2.2}) to obtain
\begin{align}
\label{eq2.3}
I_{\phi^{j}_{i}}&=\frac{8}{\omega_{i} \Delta \vec{v}^j}\int_{R^3}\int_{R^3} 
f(t,\vec{x},\vec{v}) f(t,\vec{x},\vec{v}_{1}) \frac{|\vec{g}|}{2} 
\int_{0}^{2\pi} \int_{0}^{b_{\ast}} (\phi^{j}_{i}(\vec{v}')+\phi^{j}_{i}(\vec{v}'_{1})-\phi^{j}_{i}(\vec{v})
-\phi^{j}_{i}(\vec{v}_{1})) b\, db\, d\varepsilon\, d\vec{v}_{1}\, d\vec{v} \nonumber \\
 &= \frac{8}{\omega_{i} \Delta \vec{v}^j}\int_{R^3}\int_{R^3} f(t,\vec{x},\vec{v}) f(t,\vec{x},\vec{v}_{1})
A(\vec{v},\vec{v}_{1};\phi^{j}_{i}) d\vec{v}_{1}\, d\vec{v},
\end{align}
where  
\begin{equation}
\label{eq2.4}
A(\vec{v},\vec{v}_{1};\phi^{j}_{i})= \frac{|\vec{g}|}{2} \int_{0}^{2\pi} \int_{0}^{b_{\ast}} 
(\phi^{j}_{i}(\vec{v}')+\phi^{j}_{i}(\vec{v}'_{1})-\phi^{j}_{i}(\vec{v})-\phi^{j}_{i}(\vec{v}_{1})) b\, db\, d\varepsilon \, .
\end{equation}
We notice that because $A(\vec{v},\vec{v}_{1};\phi^{j}_{i})$ is independent of time, it can 
be pre-computed and stored to be used in many individual simulations as long as the velocity 
discretization is the same. Integrals in (\ref{eq2.4}) can 
be computed with good accuracy for an arbitrary potential. 

The form (\ref{eq2.3}), (\ref{eq2.4}) 
of the discrete collision operator was first used by Pareschi and Perthame in \cite{PareschiPerthame1996} 
to achieve efficiency in a spectral Fourier discretization of the Boltzmann equation. 
In \cite{PareschiRusso2000,Pareschi2003,MouhotParechiRey2012} explicit formulas were developed for the 
components of the Fourier discretization of the collision kernel for hard spheres and Maxwell 
molecules. This form of the collision operator was also used in connection to the method of moments. 
In \cite{Bobylev1997} the form (\ref{eq2.3}), (\ref{eq2.4}) was used to develop differential 
estimates on even moments of the solution. A similar formalism is presented in detail in 
\cite{Struchtrup2005} in connection to the development of macroscopic approximations 
to the Boltzmann equation. In particular, in \cite{Struchtrup2005} expressions for collision kernels 
corresponding to globally polynomial moments are obtained in closed form 
for Maxwell molecules. Expressions for hard spheres are presented in \cite{FoxVedula2010}. 
In \cite{GreenVedula2011} and \cite{FoxVedula2010} the latter formulas are used in the context 
of Lattice-Boltzmann method to construct a closure that is based on the full Boltzmann 
collision operator. Also, in \cite{GuptaTorrilhon2012} a general algorithm is proposed to 
systematically develop values for moments of the collision operator for any collision potential. 
Most recently, the symmetric form of the collision operator was used in simulations of full 
Boltzmann equation in \cite{MouhotParechiRey2012, GhiroldiGibelli2012}. However, 
a very similar form of the collision operator appears in \cite{Aristov2001,Tcher2003}.
In particular, in \cite{Tcher2003} a formalism of Dirac delta-functions is employed in the context 
of a discrete ordinate approximation of the collision integral. We argue that the Galerkin velocity 
approach presented here can be generalized to obtain the approach of \cite{Tcher2003} by selecting appropriate 
trial ad test spaces and taking appropriate limits. 

The following properties of $A(\vec{v},\vec{v}_{1};\phi^{j}_{i})$ will be employed in our numerical method.
\begin{lem}
\label{thm2.1}
Let operator $A(\vec{v},\vec{v}_{1};\phi^{j}_{i})$ be defined by (\ref{eq2.4}) with all gas particles having
the same mass and the potential of the particles interaction being spherically symmetric. 
Then $A(\vec{v},\vec{v}_{1};\phi^{j}_{i})$ is symmetric with respect to $\vec{v}$ and $\vec{v}_{1}$, that is 
\begin{equation}
\label{eq2.5}
A(\vec{v},\vec{v}_{1};\phi^{j}_{i}) = A(\vec{v}_{1},\vec{v};\phi^{j}_{i}), \qquad \forall \vec{v},\vec{v}_{1}\in \R^3.
\end{equation} 
Also,
\begin{equation}
\label{eq2.6}
A(\vec{v},\vec{v};\phi^{j}_{i}) = 0, \qquad \forall \vec{v} \in \R^3.
\end{equation} 
\end{lem}
The proof of the lemma is in \ref{AppendixA}.

Next lemma states that $A(\vec{v},\vec{v}_{1};\phi^{j}_{i})$ is invariant with 
respect to a shift in velocity space.  
\begin{lem}
\label{thm2.2} Let operator $A(\vec{v},\vec{v}_{1};\phi^{j}_{i})$ be defined by (\ref{eq2.4}) and let 
the potential of molecular interaction be dependent only on the distance between the particles. Then $\forall\xi\in \R^3$
\begin{equation}
\label{eq2.7}
A(\vec{v}+\vec{\xi},\vec{v}_{1}+\vec{\xi};\phi^{j}_{i}(\vec{v}-\vec{\xi}))=
A(\vec{v},\vec{v}_{1};\phi^{j}_{i}) \, .
\end{equation}
\end{lem}
The proof of the lemma is in \ref{AppendixA}.

We notice that Lemma~\ref{thm2.2} allows to significantly 
reduce the required memory storage for operator $A(\vec{v},\vec{v}_{1};\phi^{j}_{i})$ in the case 
of a uniform rectangular elements and the same basis functions on each element. In this case, 
information about  $A(\vec{v},\vec{v}_{1};\phi^{j}_{i})$ needs to be stored for a single cell only. 
Values of $A(\vec{v},\vec{v}_{1};\phi^{j}_{i})$ for the rest of the cells may be restored using 
its invariance with respect to a constant shift. Of course, strictly speaking, this can only be done 
on an infinite partition of the entire velocity space. However, one can still successfully apply the 
invariance property on finite partitions provided that the support of the solution is well contained 
inside the velocity domain.

The next lemma is a generalization of Lemma~\ref{thm2.2} in the sense that it allows for more 
general transformations of the velocity space. To formulate the lemma, we need to 
recall the following definition. Consider the Euclidean space of vectors 
$\R^{3}$ with the norm defined the usual way $|\vec{v}|=\sqrt{\vec{v}\cdot\vec{v}}=\sqrt{u^2+v^2+w^2}$. 
A linear operator $S:\R^3\to \R^{3}$ is called a linear isometry if for any vector $\vec{v}\in \R^{3}$, 
\begin{equation*}
| S\vec{v} | = |\vec{v}|.
\end{equation*} 
Thus a linear isometry maps a vector into a vector of equal length. Note that this means that an 
isometry conserves distance between any two points. It follows, in particular, that an isometry 
over the entire $\R^{3}$ will transform lines into lines and spheres into spheres of the same 
radius. It will also preserve angles between lines. The most useful examples of isometries for 
us will be translations, rotations and reflections of $\R^3$. We will show next that 
operator $A(\vec{v},\vec{v}_{1};\phi^{j}_{i})$ is invariant under the action of an 
isometry, as is expressed by the next theorem. 
\begin{lem}
\label{thm2.3} Let operator $A(\vec{v},\vec{v}_{1};\phi^{j}_{i})$ be defined by (\ref{eq2.4}) and 
the potential of molecular interaction be dependent only on the distance between the particles. Let $S:\R^3\to \R^{3}$ 
be a linear isometry of $\R^{3}$. Then 
\begin{equation}
\label{eq2.8}
A(S\vec{v},S\vec{v}_{1};\phi^{j}_{i}(S^{-1}\vec{u}))=
A(\vec{v},\vec{v}_{1};\phi^{j}_{i})
\end{equation}
\end{lem}
The proof of the lemma is in \ref{AppendixA}.

\section{Discrete velocity form of the collision integral}
\label{secN5}

The numerical approximation of (\ref{eq2.3}) follows by replacing the velocity distribution 
function with the DG approximation (\ref{eq1.2}) and the integrals in (\ref{eq2.3}) with Gauss 
quadratures using the nodes $\vec{v}^{j}_{p}$. The resulting approximation of the collision operator 
takes the form
\begin{align}
\label{eq4.1}
I_{\phi^{j}_{i}}&=\frac{8}{\omega_{i} \Delta \vec{v}^j}
\sum_{j^{\ast}=1}^{M}\sum_{i^{\ast}=1}^{s}\sum_{j'=1}^{M}\sum_{i'=1}^{s}
f_{i^{\ast};j^{\ast}}(t,\vec{x})f_{i';j'}(t,\vec{x}) A^{j^{\ast}j'j}_{i^{\ast}i' i}\, . 
\end{align}
Here the quantities 
\begin{equation}
\label{eq4.2}
A^{j^{\ast}j'j}_{i^{\ast}i' i}=\frac{\Delta 
\vec{v}^{j^{\ast}}\omega_{i^{\ast}}}{8}\frac{\Delta \vec{v}^{j'}\omega_{i'}}{8} 
A(\vec{v}^{j^{\ast}}_{i^{\ast}},\vec{v}^{j'}_{i'};\phi^{j}_{i})
\end{equation}
are independent of time and are computed only once for each DG velocity discretization 
using adaptive quadratures based on Simpson's rule. It should be noted, however, that 
$A(\vec{v},\vec{v}_{1};\phi^{j}_{i})$ is not a locally supported function. In particular, 
if one of the vectors $\vec{v}$ or $\vec{v}_{1}$ falls in the support of $\phi^{j}_{i}$ 
then $A(\vec{v},\vec{v}_{1};\phi^{j}_{i})=O(|\vec{g}|)$, as $|\vec{g}| \to \infty$. If 
neither $\vec{v}$ nor $\vec{v}_{1}$ is in the support of $\phi^{j}_{i}$, 
then $A(\vec{v},\vec{v}_{1};\phi^{j}_{i})$ is decreasing as $|\vec{g}| \to \infty$. 
However, it does not equal zero, no matter how large is $|\vec{g}|$. Because of 
this, additional considerations must be employed in order to limit the number 
of entries $A^{j^{\ast}j'j}_{i^{\ast}i' i}$ that will be stored. In our approach, 
two strategies are employed. The first strategy eliminates all values of 
$A^{j^{\ast}j'j}_{i^{\ast}i' i}$ whose magnitudes fall below the levels of 
expected errors in the adaptive quadratures. The second strategy eliminates 
values of $A^{j^{\ast}j'j}_{i^{\ast}i' i}$ that correspond to pairs 
$\vec{v}^{j^{\ast}}_{i^{\ast}}$ and $\vec{v}^{j'}_{i'}$ for which 
$|\vec{v}^{j^{\ast}}_{i^{\ast}} - \vec{v}^{j'}_{i'}|$ is greater than some 
specified diameter. While the rationale for the first strategy is self-explanatory, 
the second strategy can be justified by the fact that solutions to the Boltzmann equations 
decrease rapidly at infinity. They can generally be assumed to be zero outside of a ball 
of the diameter of several thermal velocities with the center at the stream velocity. 
Because in most cases, thermal velocity can be estimated without knowing much about the 
final solution, one can limit $A^{j^{\ast}j'j}_{i^{\ast}i' i}$ to only those pairs of 
$\vec{v}^{j^{\ast}}_{i^{\ast}}$ and $\vec{v}^{j'}_{i'}$ that are contained in such ball. 

Table~\ref{tab01} illustrates the 
growth rate of the number of non-zero entries in $A^{j^{\ast}j'j}_{i^{\ast}i' i}$ 
for a single basis function $\phi^{j}_{i}$. The dimensionless velocity domain is the cube with 
sides $[-3,3]$ in each dimension. Two cases of DG basis are considered. 
The first one corresponds to DG approximations by constants given by $s_{u}=s_{v}=s_{w}=1$. 
The second one corresponds to DG approximations by quadratic polynomials given by 
$s_{u}=s_{v}=s_{w}=3$. Different number of velocity cells were used. Velocity pairs 
separated farther than $3$ units were neglected. Threshold for adaptive quadrature error 
was set at $10^{-8}$. 

The top row of Table~\ref{tab01} corresponds to the numbers of degrees of freedom, $N$, in each 
velocity dimension. For example, if $s_{u}=s_{v}=s_{w}=3$ and the velocity grid has 
$M=5$ velocity cells in each dimension we obtain the number of degrees of freedom in each dimension to be 
$N=s_{u}M=15$. The number of non-zero 
components of $A^{j^{\ast}j'j}_{i^{\ast}i' i}$ are listed in rows two and four. 
Estimated orders of growth are shown in rows three and five. It is observed that 
the number of non-zero components grows approximately as $O(N^5)$ in both piece-wise constant and 
piece-wise quadratic cases. This is considerably smaller than the expected growth rate of 
$O(N^6)$ obtained by counting of all possible pairs of pre-collision velocities. The reduction in 
size of $A^{j^{\ast}j'j}_{i^{\ast}i' i}$ can be attributed to the locality 
of the DG basis functions. Indeed when basis functions are locally supported, 
most of pairs $\vec{v}^{j^{\ast}}_{i^{\ast}}$ and $\vec{v}^{j'}_{i'}$ produce a collision sphere 
that does not overlap with the support of the basis function $\phi^{j}_{i}$.  As a result, 
the corresponding values of $A^{j^{\ast}j'j}_{i^{\ast}i' i}$ for all such combinations are zero.

\begin{table}[htp]
\begin{tabular}{lrrrrr}
\hline
Nodes per dimension, $N$ & 9 & 15 & 21 & 27 & 33 \\
Components, $s_{u}=1$ & 11278& 143804 & 781002 & 2693240 & 7261854 \\
Order, $s_{u}=1$ &4.98 & 5.03& 4.93& 4.94 & {} \\
Components, $s_{u}=3$ & 39022 & 459455	& 2355130 & 8142006 & 21915065\\
Order, $s_{u}=3$ &4.83 & 4.86& 4.94 & 4.93 & {} \\
\hline
\end{tabular}
\caption{\label{tab01} The growth of the number of non-zero components of $A^{j^{\ast}j'j}_{i^{\ast}i' i}$ for a 
single basis function $\phi^{j}_{i}$ with respect to the numbers of velocity nodes and the estimated orders 
of growth.}
\end{table}

It is however unlikely that $O(N^5)$ can be further reduced by, say, a more elaborate 
choice of the DG basis. Each evaluation of the Boltzmann collision equation in its original 
form requires the knowledge of $f(t,\vec{x},\vec{v})$ at about $N^5$ pairs of velocity points. 
Indeed, to evaluate the collision operator one has to consider $N^3$ different selections of 
the second pre-collision velocity and to pair each selection with about 
$N^2$ combinations of impact parameters to produce on the order of $N^{5}$ 
various combinations of $\vec{v}$, $\vec{v}_{1}$, $\vec{v}'$ and $\vec{v}'_{1}$ that 
are averaged in (\ref{boltzm2}). This suggests that (\ref{eq2.3}) maintains the same information 
as the direct discretization of the full collision operator. Because of this, it is unlikely that 
the complexity $O(N^5)$ can be further reduced without restricting the collision operator itself. 

It can be seen that the number of non-zero components is larger in the case of $s_{u}=3$ than in 
the case of $s_{u}=1$ for the same number of degrees of freedom, $N$. This is explained by the fact 
that supports of the basis functions are larger in the case of $s=3$. Therefore, more pairs of 
velocities produce non-negligible integrals.

By multiplying the amount of storage required for one basis function, $O(N^5)$, by the total number 
of basis functions, $N^3$, one estimates the total amount of storage for $A^{j^{\ast}j'j}_{i^{\ast}i' i}$ 
to grow as $O(N^8)$. However, this number can be reduces back to $O(N^5)$ by using uniform partitions 
and Lemmas~\ref{thm2.2} and \ref{thm2.3}. Indeed, by identifying ways to map combinations of triples 
$\vec{v}^{j^{\ast}}_{i^{\ast}}$, $\vec{v}^{j'}_{i'}$ and $\phi^{j}_{i}$ into each other via 
linear isometries, a small subgroup of unique records in $A^{j^{\ast}j'j}_{i^{\ast}i' i}$ 
can be determined. Only this subgroup needs to be computed and stored. Records for the rest of 
the triples can be restored using (\ref{eq2.7}) and (\ref{eq2.8}). 

In the simulations presented in this paper, the velocity domain was partitioned into uniform rectangular 
parallelepipeds and the same Lagrange basis functions were used on each element. One can notice that in this case, all 
cells can be obtained from a single cell by a constant shift. One also notices that basis functions and nodes 
can be obtained from the basis functions and the nodes of that selected cell using the substitution described in  
Lemma~\ref{thm2.2}. It follows then that records can be restored from the records of the canonical 
cell using (\ref{eq2.7}) and that only records of the canonical cell need to be stored. Of course, in the case of 
a finite partition some shifts will produce values of the velocity that are not on the grid. 
However, if the support of the solution is well contained inside the domain, such values can 
be ignored in the summation of (\ref{eq4.1}). Finally, if the nodes and basis functions have 
rotational symmetries within the element, more isometries can be considered and 
the storage for $A^{j^{\ast}j'j}_{i^{\ast}i' i}$ can be further reduced. However, it will 
still be proportional to $N^5$ even if more symmetries are found. 

Because the components of $A^{j^{\ast}j'j}_{i^{\ast}i' i}$ are independent of each other, their 
evaluation can be parallelized using hundreds of processors. In the simulations presented in 
this paper up to $320$ processors were used with MPI parallelization algorithms. While the current 
implementation allows, in principal, to evaluate up to $N=100$ degrees of freedom in one dimension 
and for larger values of $s_{u}$, this has not been done due to the limited 
available computer resources. 

Times to compute (\ref{eq4.1}) for a single time step on a single 2.3 MHz processor are presented in 
Table~{\ref{tab02}}. One can notice that the computation time grows as $O(N^8)$. This is not at all surprising since the 
the $O(N^5)$ operations for evaluation of collision operator need to be repeated at $O(N^3)$ velocity nodal 
points. The computational time grows at slightly higher rate for $s=1$
than for $s=3$. This can be explained by the fact that in the case of
$s=1$, $A^{j^{\ast}j'j}_{i^{\ast}i' i}$ is only stored at one node in 
the center of the grid $s=1$ as compared to 27 nodes in the case of
$s=3$. The rest of the values are restored by Lemma~\ref{thm2.2} in both cases.
However, in the case of $s=1$ the application of Lemma~\ref{thm2.2} 
involves significantly more memory copying. Because memory copying is 
expensive, this causes computational time for $s=1$ grow at a faster rate. 
It is however believed that summation routines can be re-formulated 
so as to minimize the memory copying, see e.g. \cite{cacheoblivious}. However, even 
with the efficient summation the growth rate of $O(N^8)$ is 
believed to be intrinsic to the method and therefore is expected 
to quickly saturate the computational resources. 
However, it will be seen in the next section that the method yields 
reasonable calculation times on a single processor for values of $N$ 
up to $33$. Because of the fast growth of computational time 
it is expected that somewhere from $10^3$ to $10^4$ processors will be required 
to reach the value of $N=100$. Because the method
involves very little data exchange, we expect that 
parallelization of the algorithm will be very efficient. However, 
the main savings are expected from the construction of 
efficient Galerkin basis so as to minimize the total number 
of degrees of freedom while maintaining accuracy. Construction 
of such approximations will be the topic of the authors' future work.

\begin{table}[htp]
\begin{tabular}{lrrr}
\hline
Number of nodes, $N$ & 9 & 15 & 21 \\
Processor time in seconds, $s_{u}=1$ & 0.14 & 8.63 & 134.07 {} \\
Order, $s_{u}=1$ &8.05 & 8.15 & {} \\
Processor time in seconds, $s_{u}=3$ & 0.65 & 35.17 &486.82 \\
Order, $s_{u}=3$ &7.81 & 7.81 & {}\\
\hline
\end{tabular}
\caption{\label{tab02} Time in seconds to advance one temporal step.}
\end{table}

\section{Spatially homogeneous relaxation}
\label{secN6}

To illustrate the work of the algorithm we present results of
simulations of relaxation of monoatomic gas from perturbed states.
Two problems are considered: relaxation of two equilibrium streams and
relaxation of a discontinuous initial stage. The
molecular collision was modelled using hard spheres potential in both 
problems. Two instances of DG discretizations were compared: approximations by 
piece-wise constants, corresponding to $s_{u}=s_{v}=s_{w}=1$, and by piece-wise 
quadratic approximations, corresponding to $s_{u}=s_{v}=s_{w}=3$. The time 
discretization in all simulations is by fifth order Adams-Bashforth method. 
The data for the Adams-Bashforth method is obtained using the fifth 
order Runge-Kutta method. 

In the first problem the initial data is the sum of two Maxwellian distributions 
with mass densities, bulk velocities and temperatures 
of $\rho_{1}=6.634$E$-6$ kg/m${}^3$,
$\vec{\bar{v}}_{1}=(967.78,0,0)$ m/s, and $T_{1}=300$ K and
$\rho_{2}=1.99$E$-5$ kg/m${}^3$,
$\vec{\bar{v}}_{2}=(322.59,0,0)$ m/s, and $T_{2}=1100$ K correspondingly. 
The spatially homogeneous relaxation is simulated for about 120 $\mu$s. 
The mean time between collisions for the steady state solution is estimated to be about 5.4~$\mu$s.
The solution appeared to reach the steady state at about 45~$\mu$s. 

\begin{figure}[tp]
  \begin{tabular}{@{}cc@{}}
  \includegraphics[height=.25\textheight]{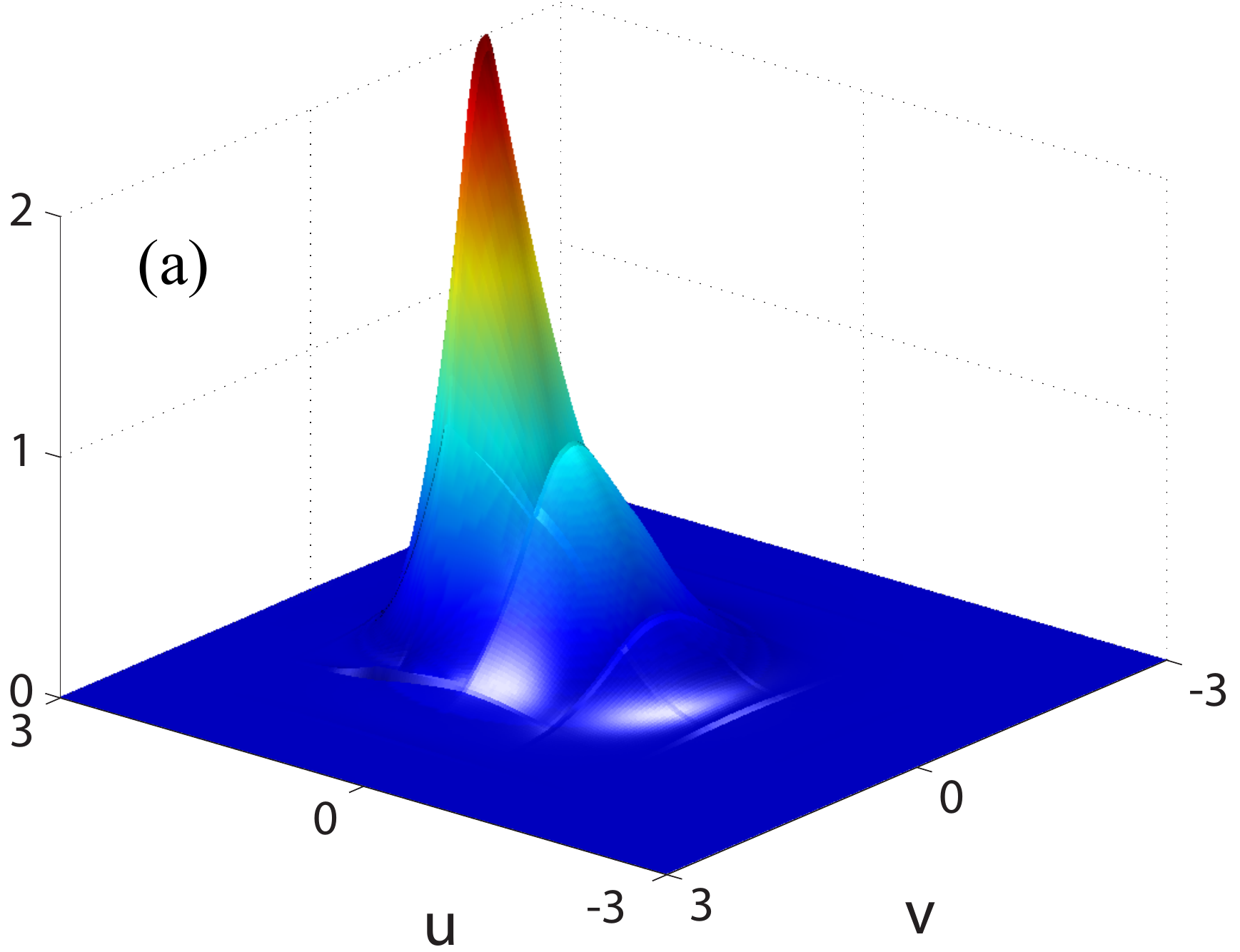}&
  \includegraphics[height=.25\textheight]{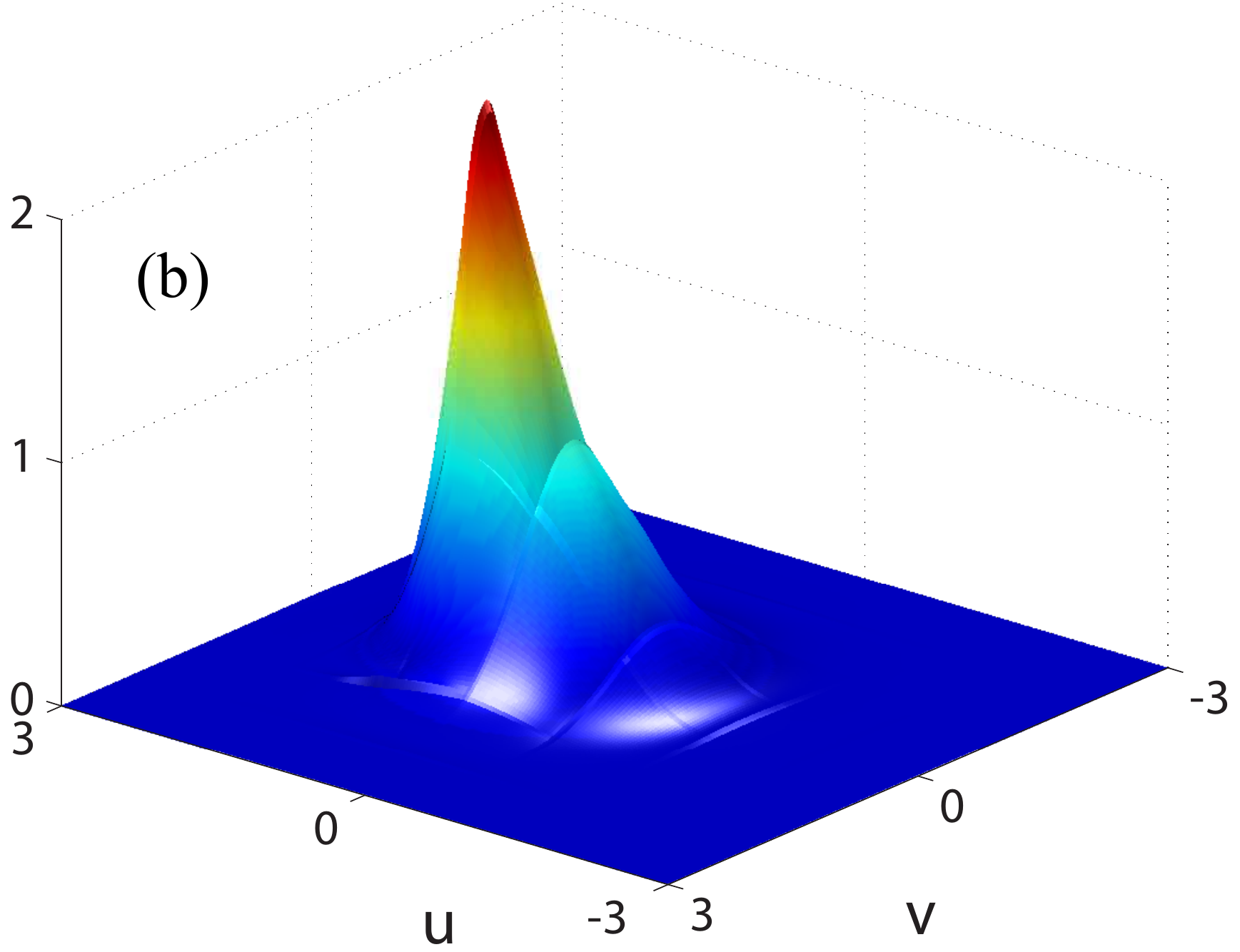}\\
  \includegraphics[height=.25\textheight]{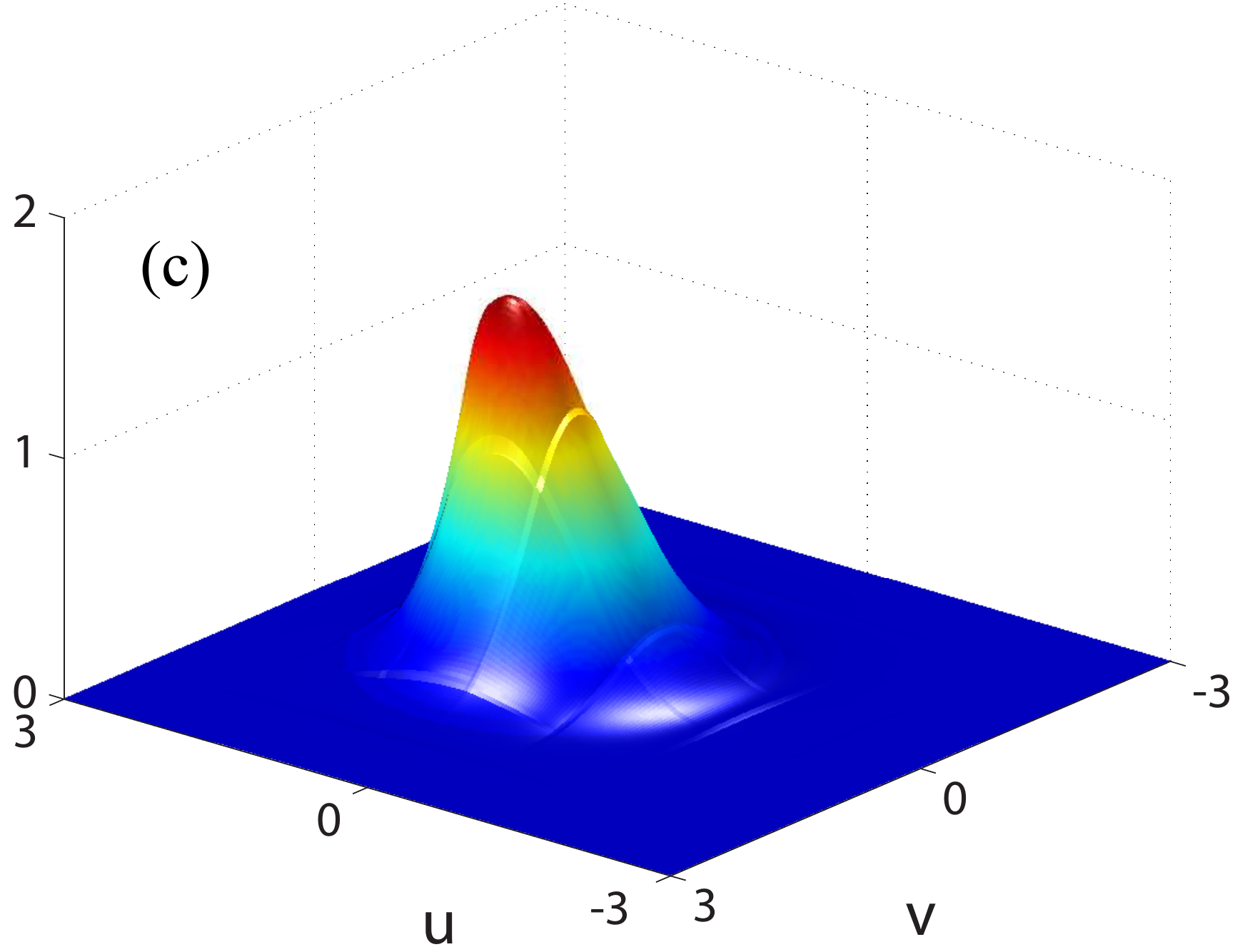}&
  \includegraphics[height=.25\textheight]{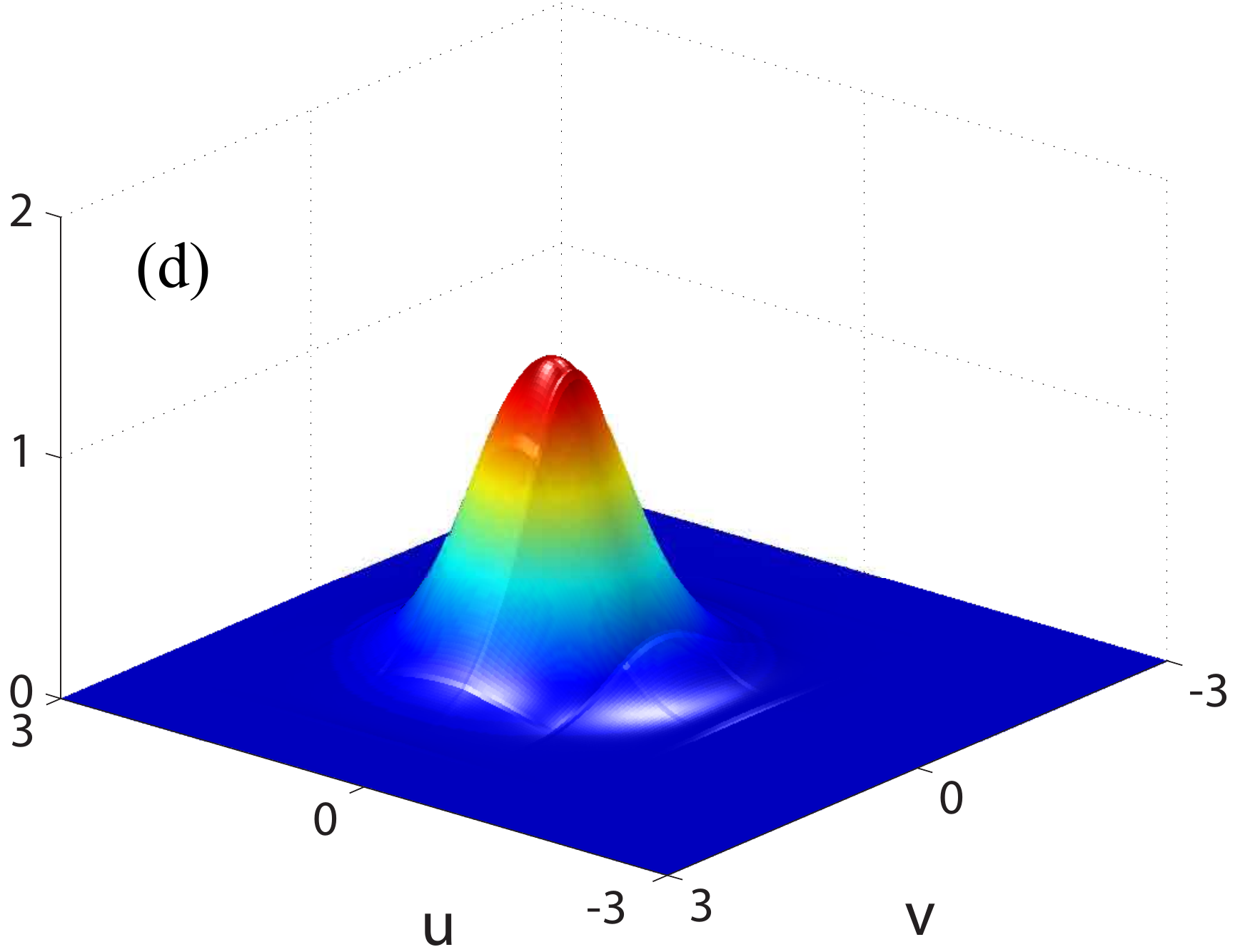}\\
  \end{tabular}
  \caption{\label{fig01} Relaxation of two Maxwellian streams using piece-wise quadratic DG approximations. 
  Sections of the solution by plane $w=0$ are shown. The solutions correspond to $s_{u}=s_{v}=s_{w}=3$ and 
  7 cells in each velocity direction. (a) $t=0$ $\mu$s, (b) $t=1.3$ $\mu$s,
  (c) $t=7.6$ $\mu$s, and (d) $t=39.2$ $\mu$s.}
\end{figure}

\begin{figure}[tp]
  \begin{tabular}{@{}cc@{}}
  \includegraphics[height=.25\textheight]{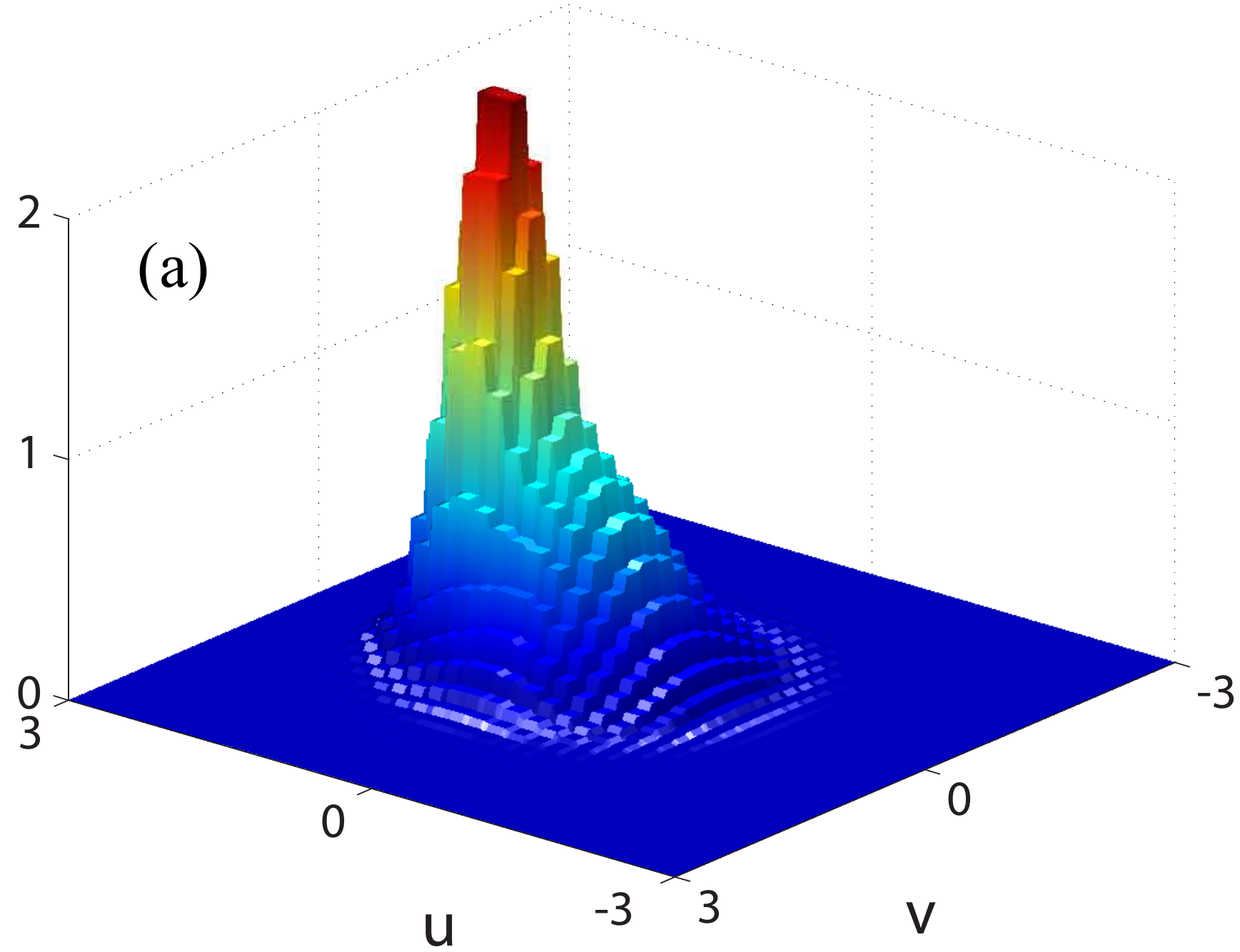}&
  \includegraphics[height=.25\textheight]{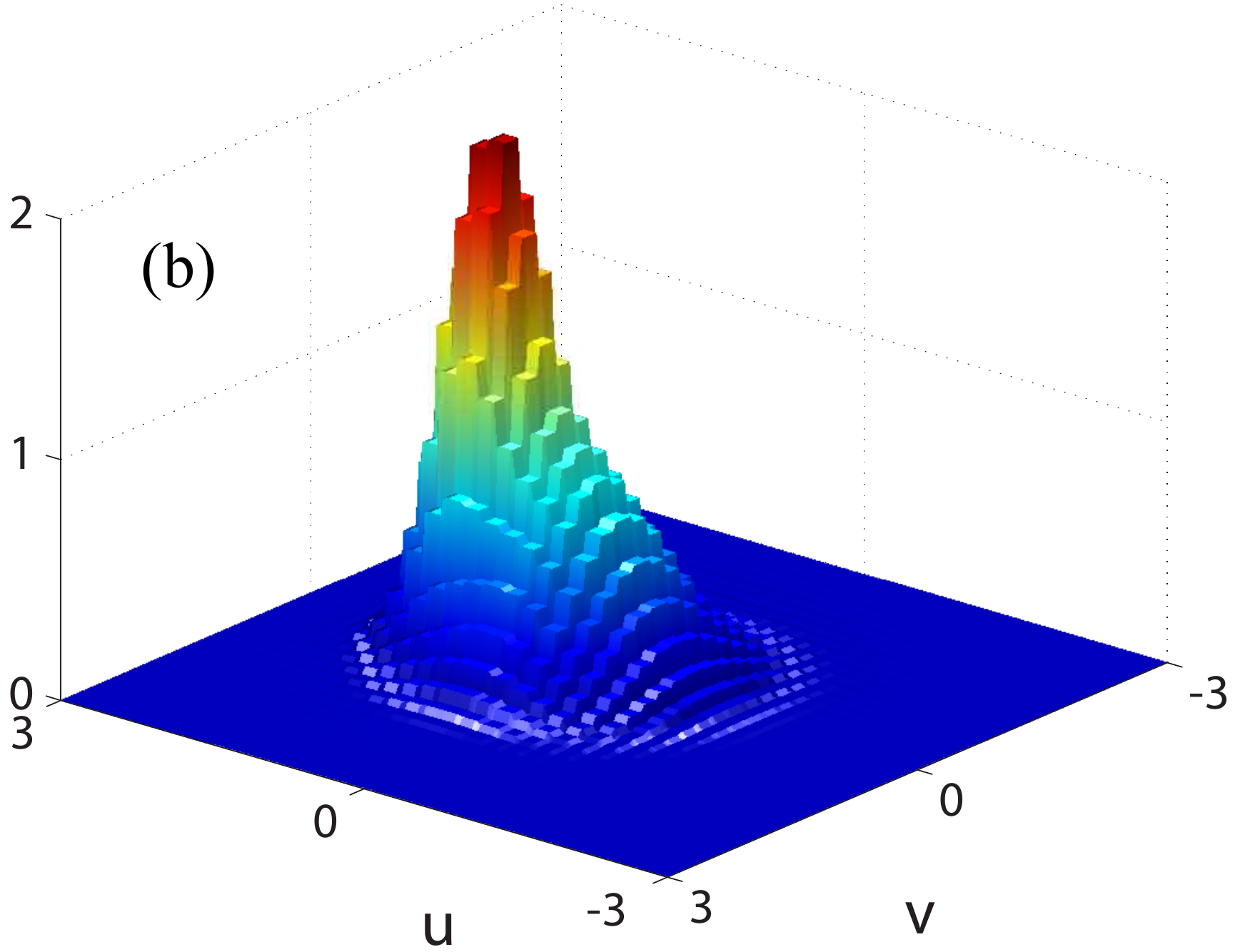}\\
  \includegraphics[height=.25\textheight]{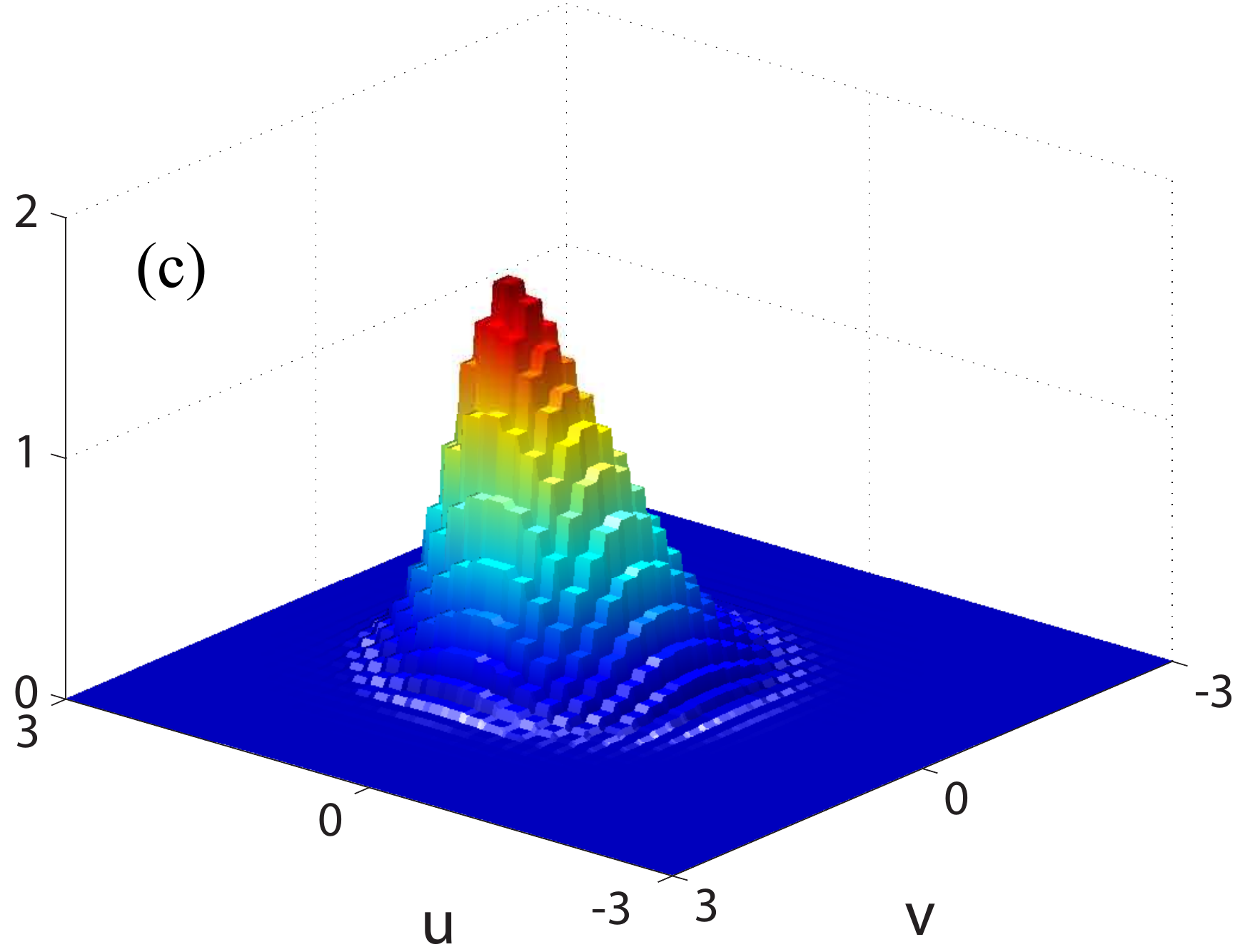}&
  \includegraphics[height=.25\textheight]{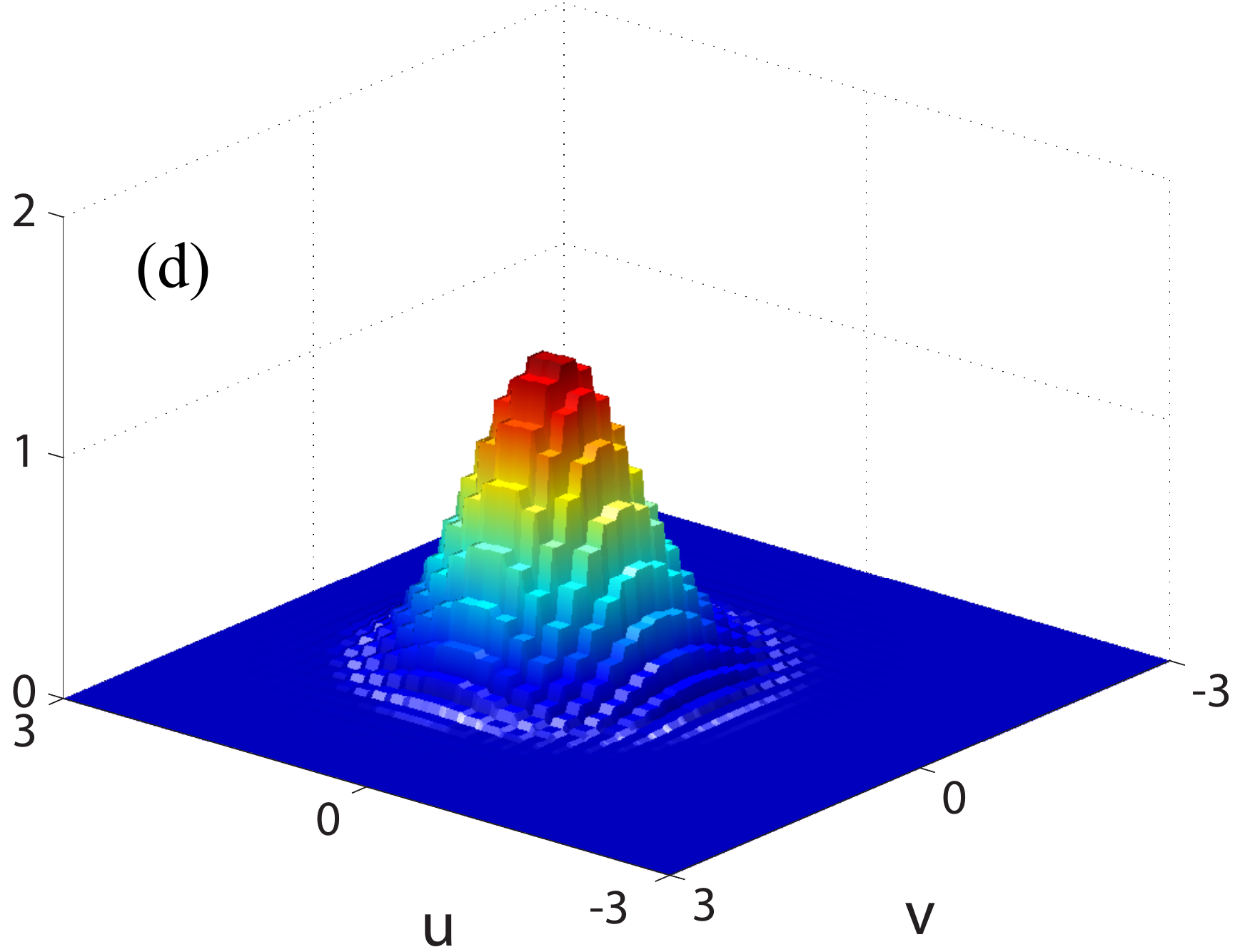}\\
  \end{tabular}
  \caption{\label{fig01a} Relaxation of two Maxwellian streams using piece-wise constant DG approximations. 
  Sections of the solution by plane $w=0$ are shown. The solutions correspond to $s_{u}=s_{v}=s_{w}=1$ and 
  33 cells in each velocity direction: (a) $t=0$ $\mu$s; (b) $t=1.3$ $\mu$s;
    (c) $t=7.6$ $\mu$s; and (d) $t=39.2$ $\mu$s.}
\end{figure}

\begin{figure}[thp]
  \begin{tabular}{@{}c@{}c@{}}
  \includegraphics[height=.285\textheight]{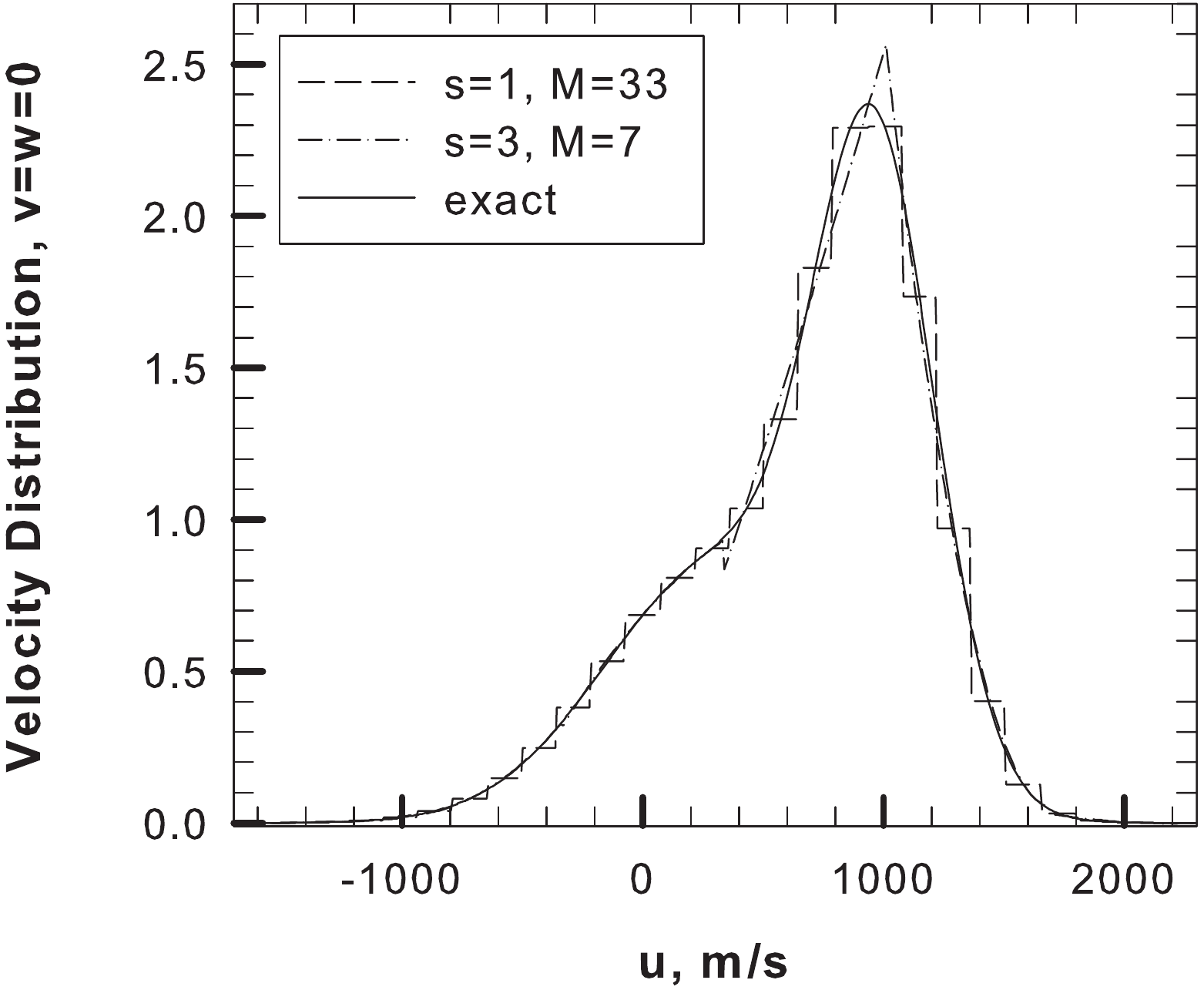}&
  \includegraphics[height=.285\textheight]{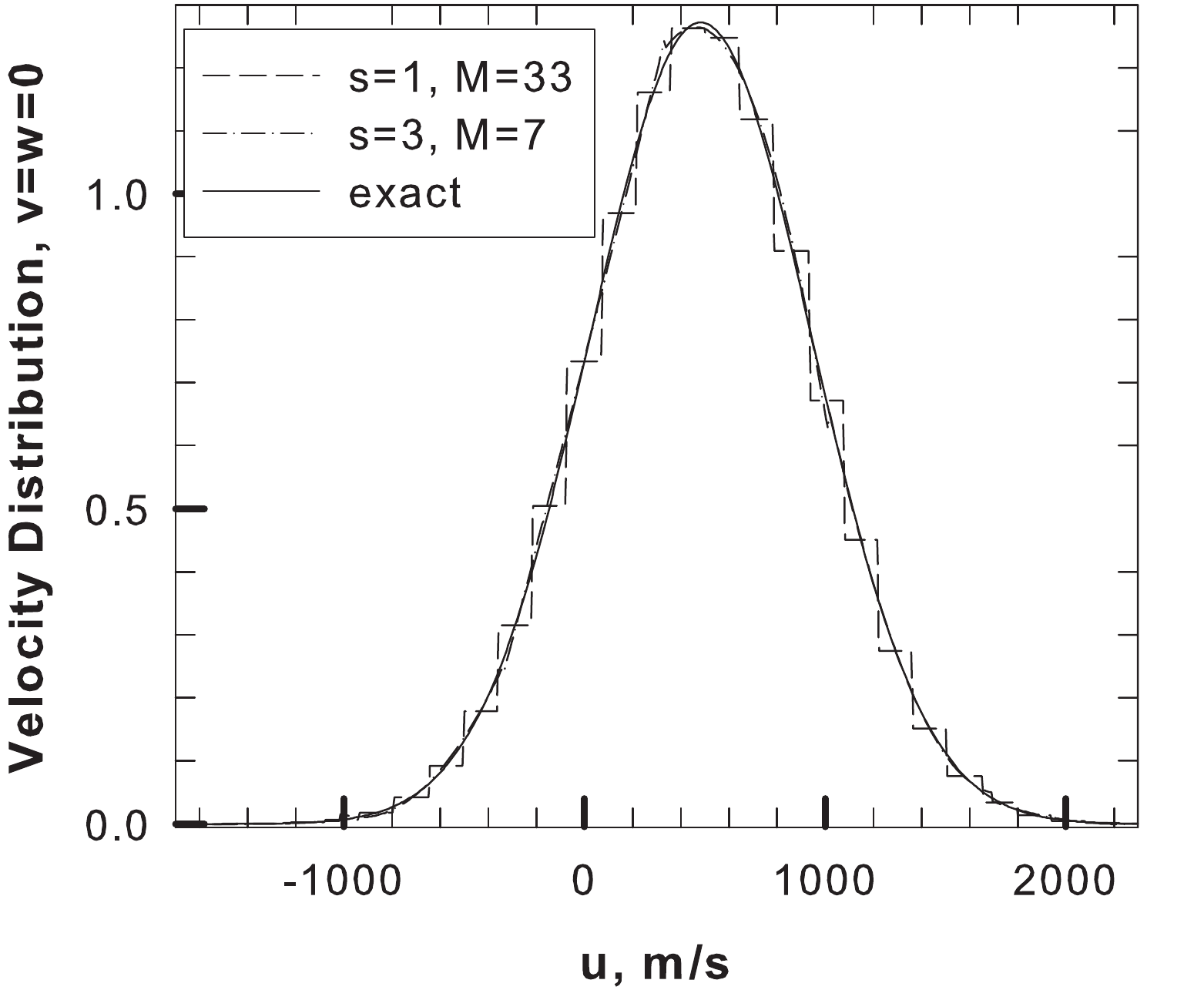}\\
  {\small (a) }&{\small (b) }
  \end{tabular}
  \caption{\label{fig03} Cross sections of the solution to 
  the problem of relaxation of two Maxwellian streams by planes $v=0$ and $w=0$. 
  DG approximations with $s_{u}=s_{v}=s_{w}=1$, $M=27$ and 
  $s_{u}=s_{v}=s_{w}=3$, $M=7$ are compared. (a) Approximations of the initial data. 
  (b) Approximations of the steady state.}
\end{figure}

In Figures~{\ref{fig01}} and \ref{fig01a} solutions to the relaxation of the two Maxwellian streams 
are shown. Simulations corresponding to $s_{u}=s_{v}=s_{w}=3$, $M=7$ are shown in 
Figure~{\ref{fig01}} and the simulation for $s_{u}=s_{v}=s_{w}=1$, $M=33$ 
in Figure~{\ref{fig01a}}. One-dimensional sections of the solution 
by planes $v=0$ and $w=0$ are shown in Figure~{\ref{fig03}}. In Figure~{\ref{fig03}}(a) 
the DG approximations of the initial data are given and in Figure~{\ref{fig03}}(b) the 
approximations of the steady state are given. Both the piece-wise constant and the 
piece-wise quadratic case appear to capture the relaxation process successfully. 

\begin{figure}[t]
  \begin{tabular}{@{}c@{}c@{}}
  \includegraphics[height=.3\textheight]{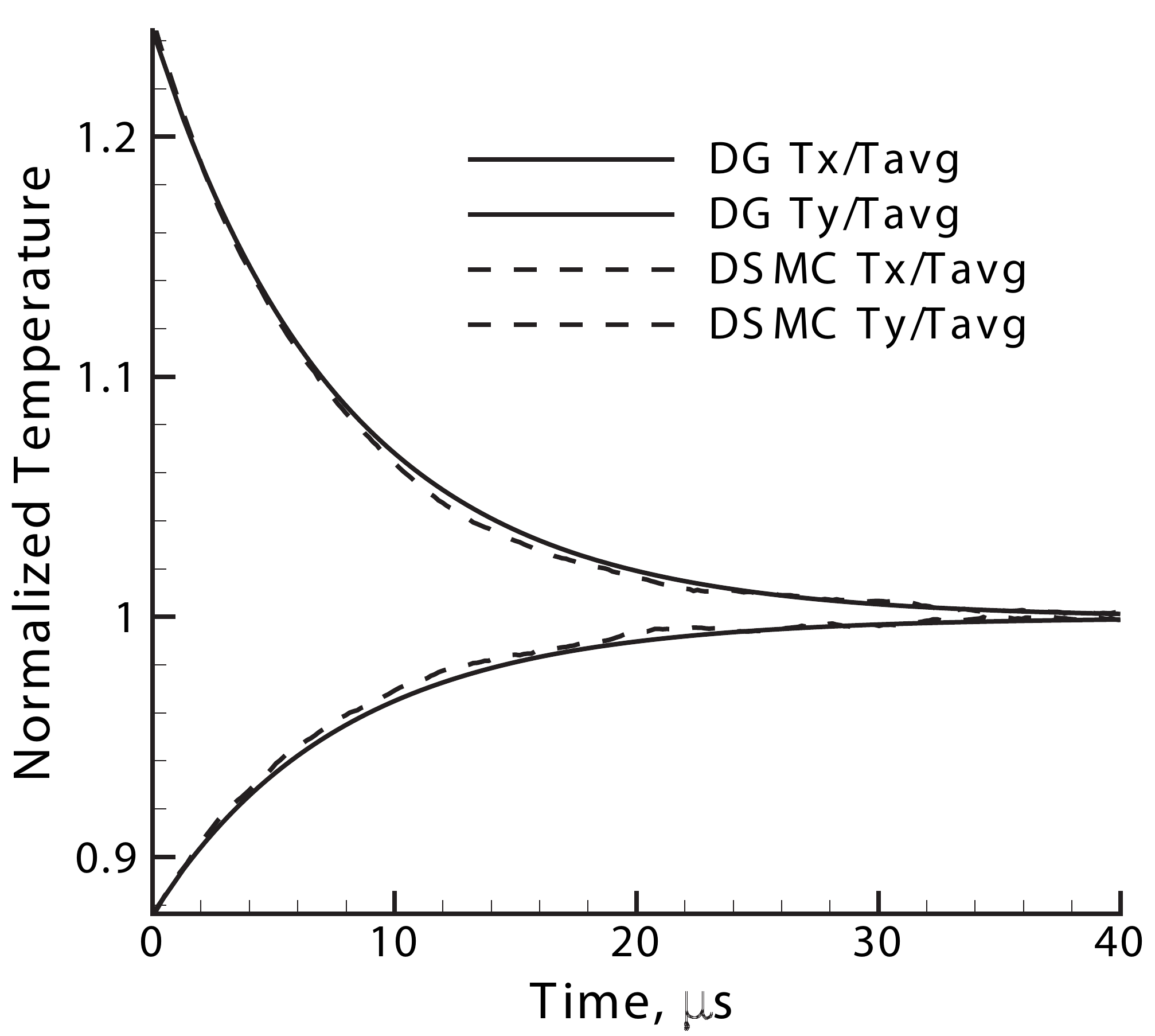}&
  \includegraphics[height=.3\textheight]{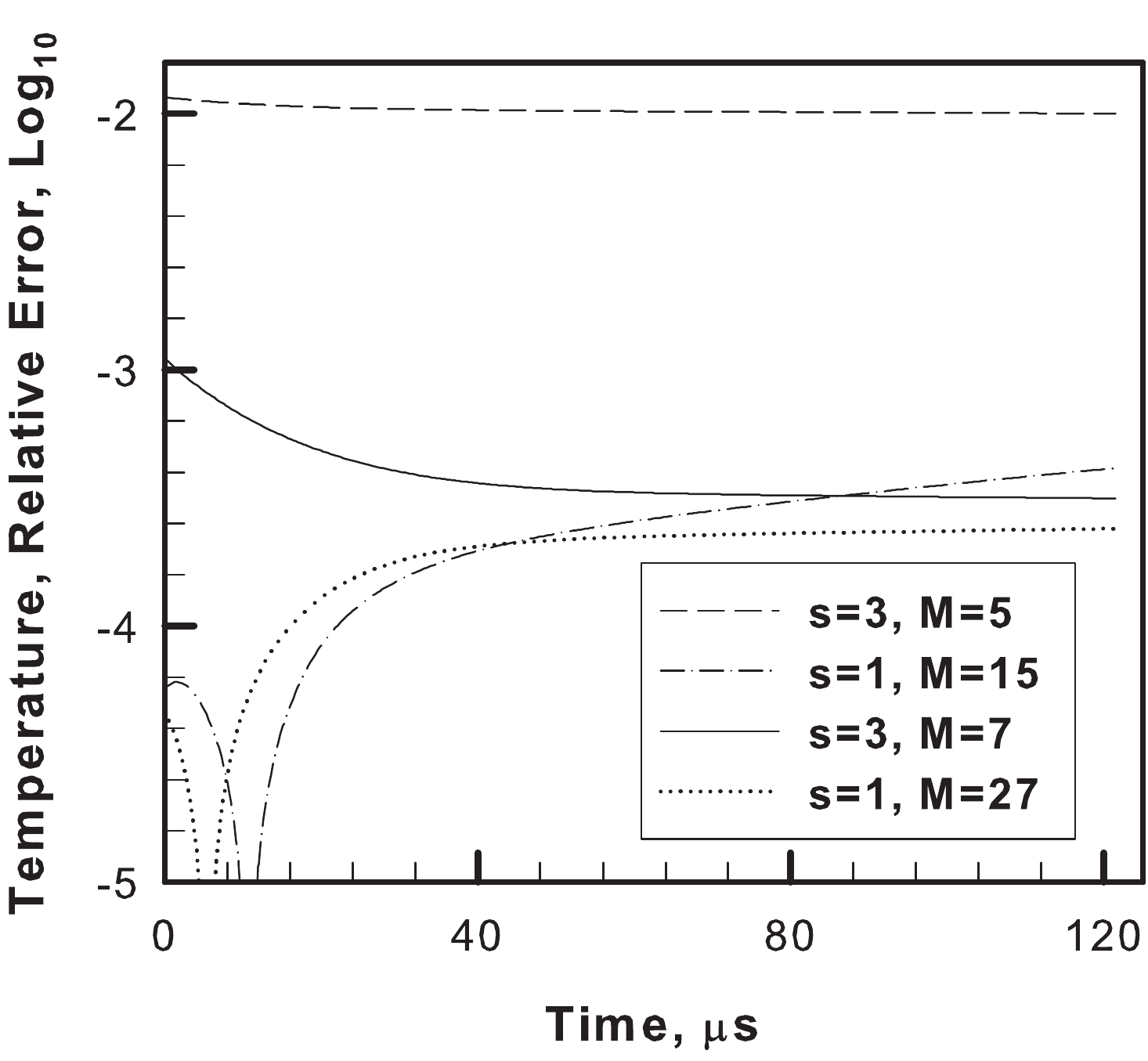}\\
  {\small (a) }&{\small (b) }
  \end{tabular}
  \caption{\label{fig02} Relaxation of two Maxwellian streams. (a) Comparison of 
  the DG solution to the DSMC solution by I. Boyd \cite{Boyd1991411}. Ratios
  $T_{x}/T_{\mathrm{avg}}$ and  $T_{y}/T_{\mathrm{av g}}$ are plotted. Here $T_{x}$ and 
  $T_{y}$ are directional temperatures and $T_{\mathrm{avg}}=T/3$, where $T$ 
  is the temperature of the exact solution. (b) Relative errors in the solutions 
  temperature. Piece-wise quadratic and piece-wise constant DG approximations are compared.
  The piece-wise constant approximations are more accurate in this problem.}
\end{figure}

DG solutions were compared with the solutions obtained by 
established DSMC solvers \cite{Boyd1991411}. In Figure~\ref{fig02}(a) the 
ratios of directional temperatures $T_{x}$ and $T_{y}$ to the average 
temperature $T_{\mathrm{avg}}=T/3$ are shown. The solution reaches 
equilibrium state at about 45 $\mu$s. The DG approximation shows 
an excellent agreement with the DSMC solution. In Figure~\ref{fig02}(b) 
relative errors in the solution temperature are 
presented for $s_{u}=s_{v}=s_{w}=3$, $M=5,7$ 
and $s_{u}=s_{v}=s_{w}=1$, $M=15,27$. In our DG approach no special 
enforcement of conservation laws is used. Rather, it is expected that similar to \cite{AGG12,Alekseenko2011}
a satisfactory conservation can be achieved by using sufficiently refined DG approximations. 
It can be observed that for $s_{u}=s_{v}=s_{w}=33$, $M=7$ the 
temperature is computed correctly within three digits of accuracy. It was 
observed also that piece-wise constant approximations perform significantly 
better in this problem. This finding is consistent with \cite{Alekseenko2011} 
where it was found that piece-wise constant DG approximations are converging faster 
than high order DG approximations on smooth solutions. It is however expected that 
high order methods will be superior on 
the solutions that are not numerically 
smooth, e.g., due to truncation errors in spatially inhomogeneous problems.

\begin{figure}[h]
  \begin{tabular}{@{}cc@{}}
  \includegraphics[height=.25\textheight]{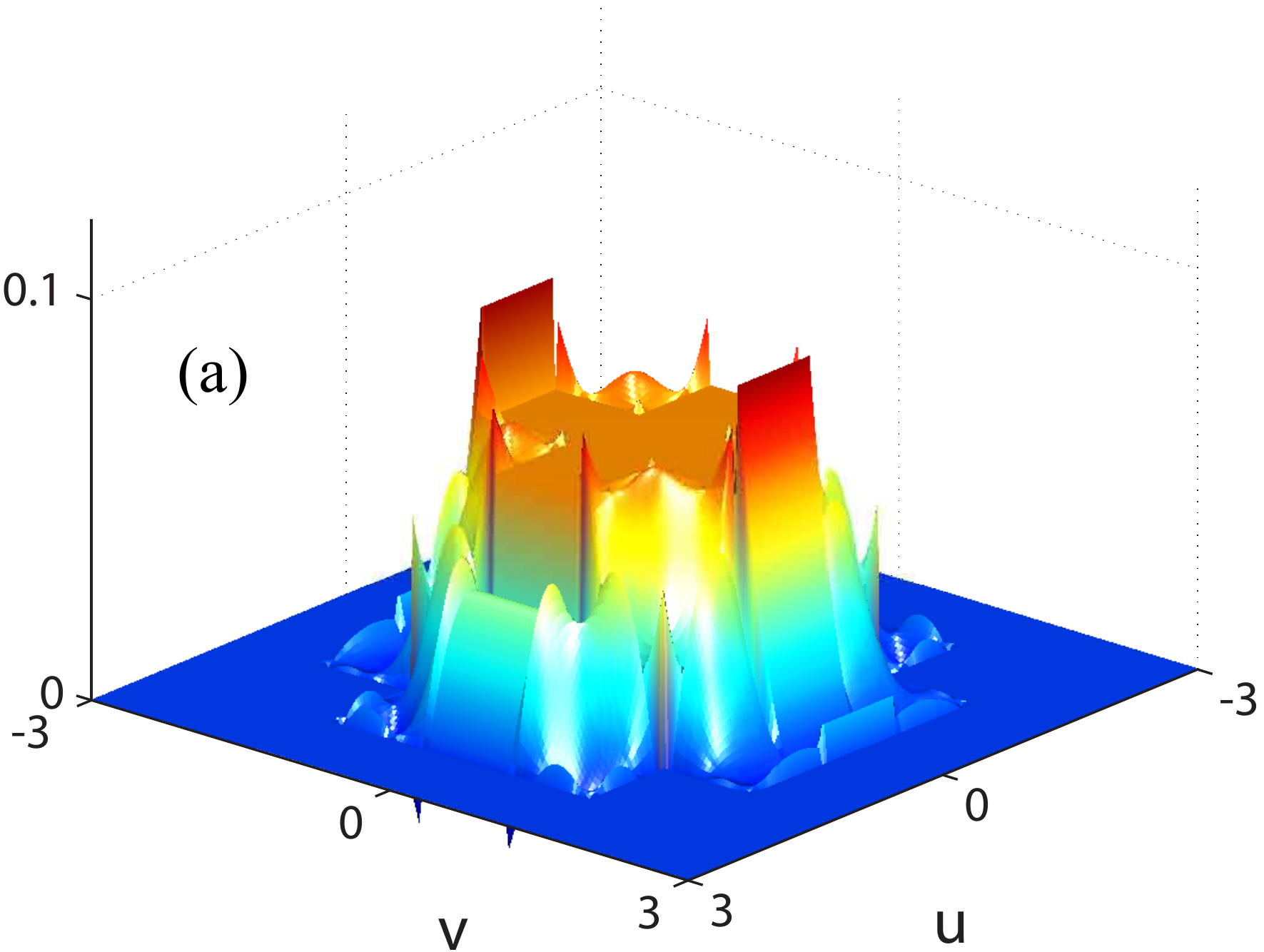}&
  \includegraphics[height=.25\textheight]{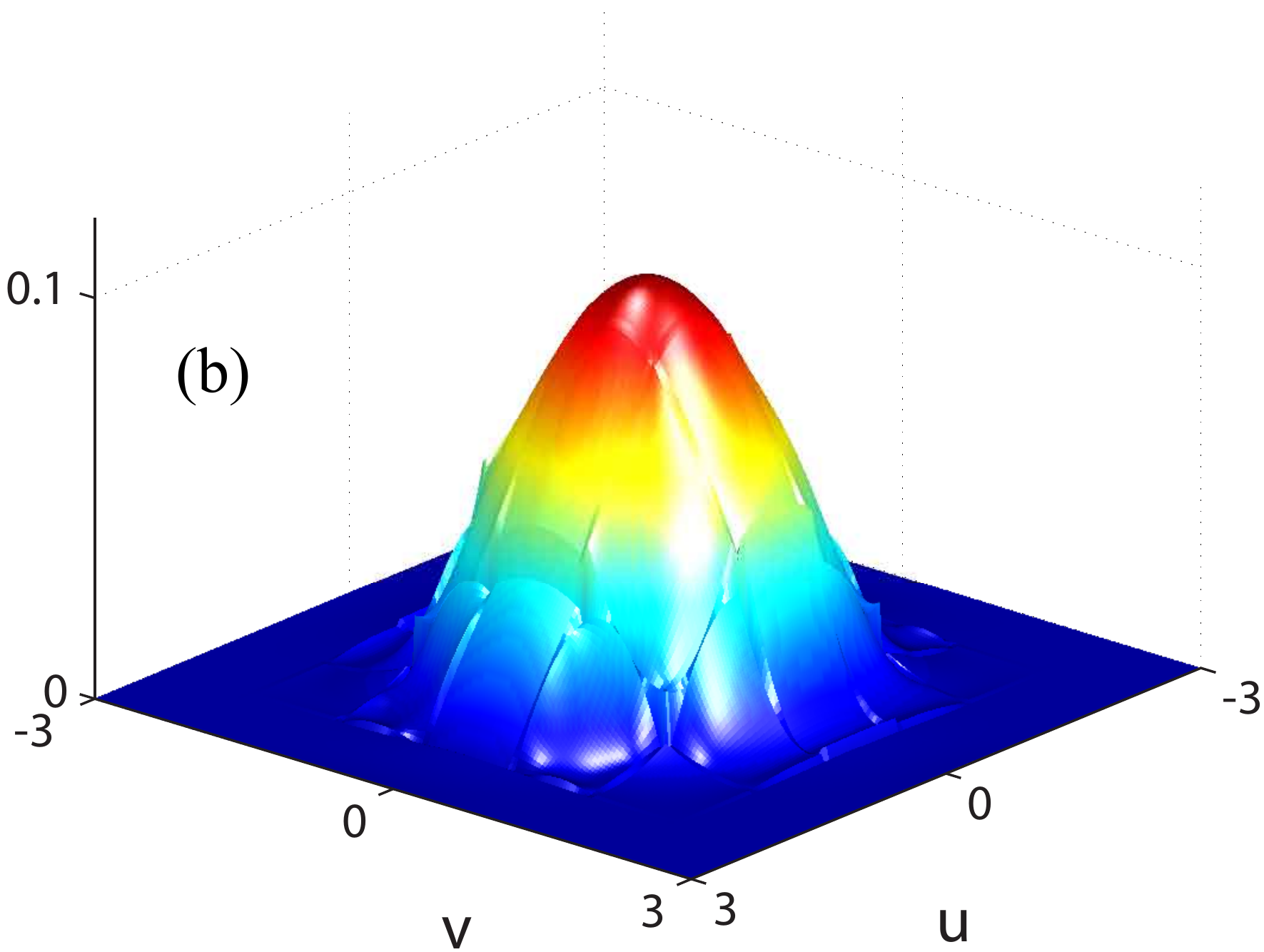}\\
  \includegraphics[height=.25\textheight]{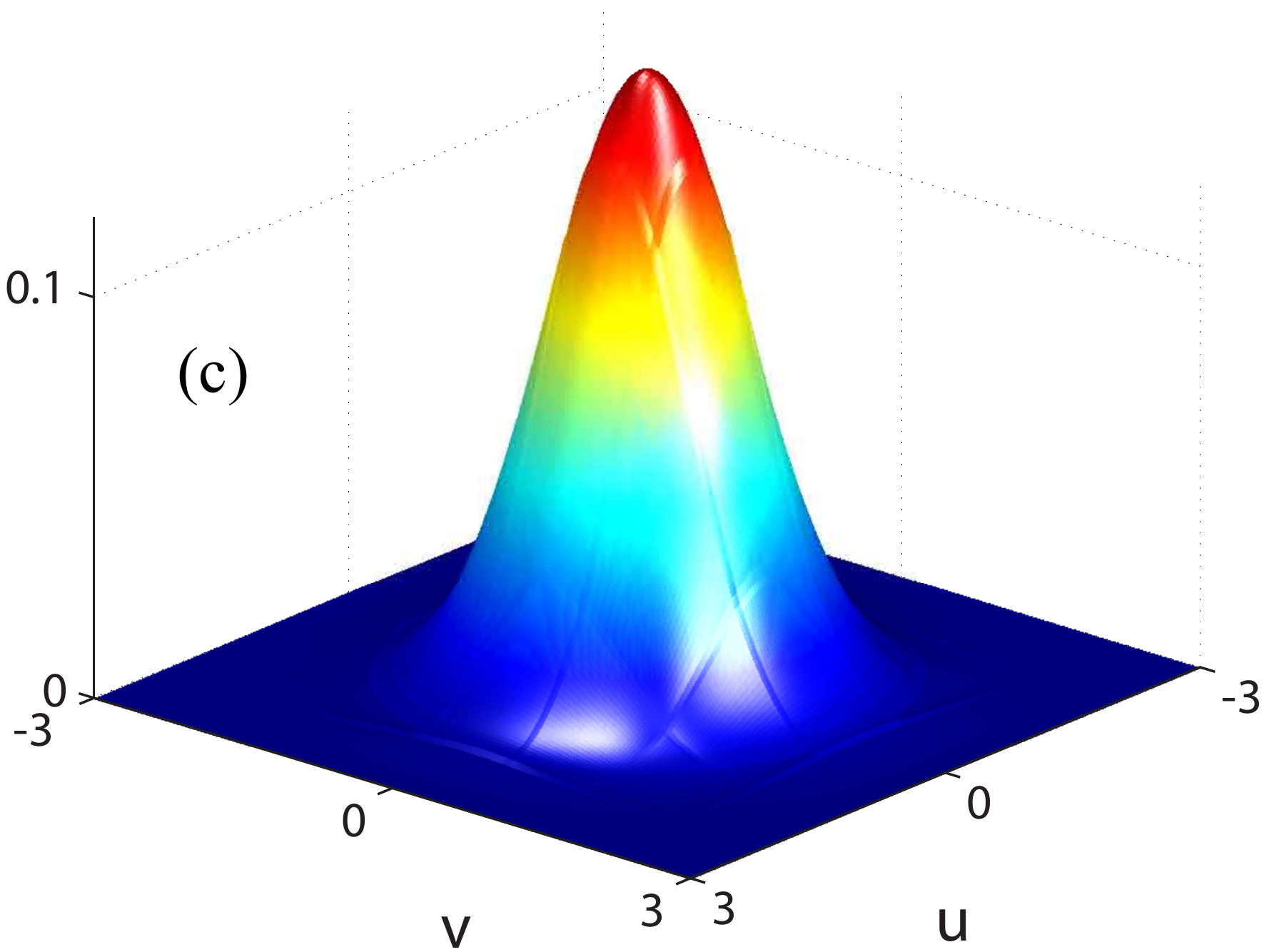}&
  \includegraphics[height=.25\textheight]{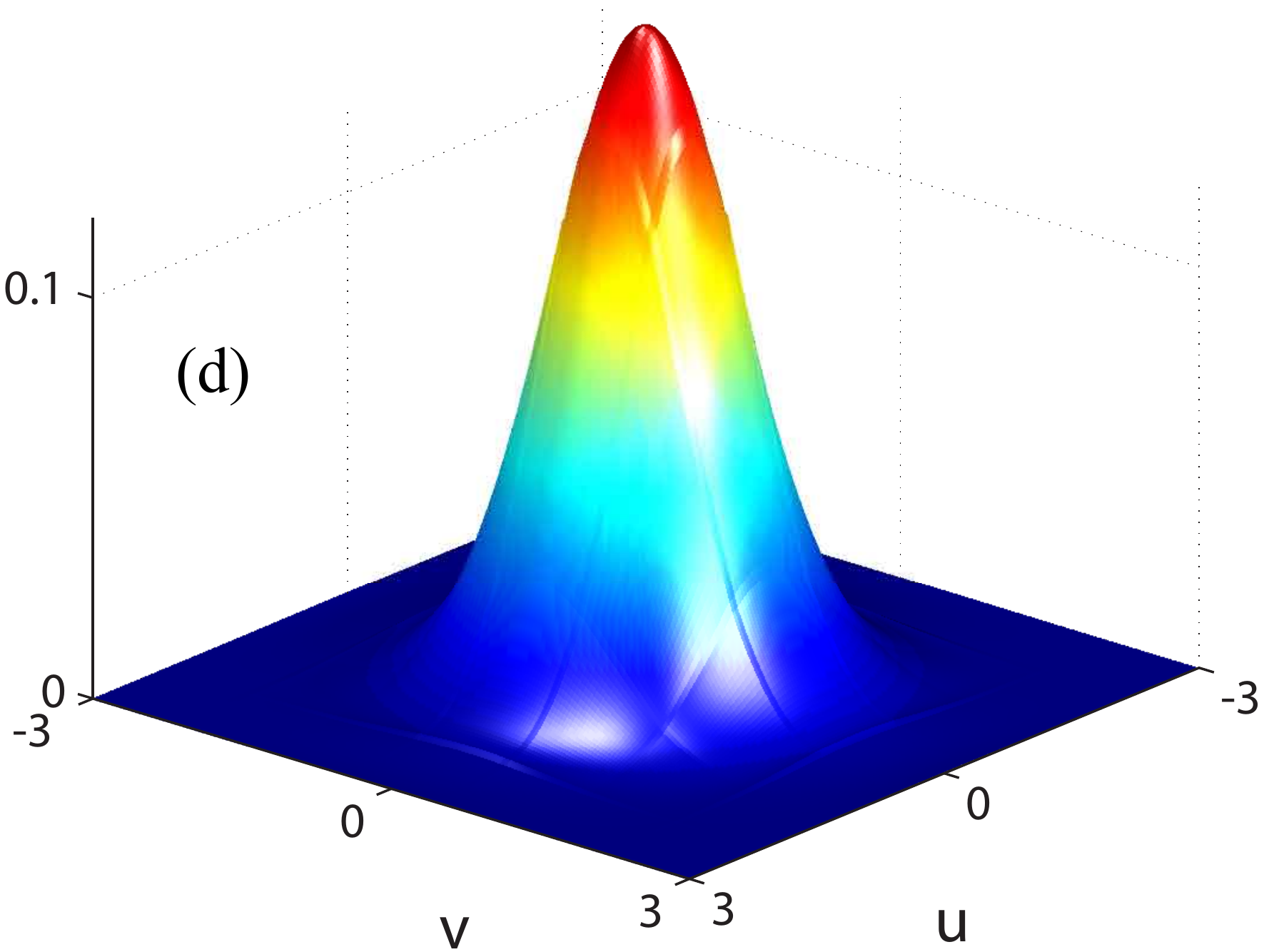}\\
  \end{tabular}
  \caption{\label{fig04} Relaxation of two artificial streams using piece-wise quadratic DG approximations. 
      Sections of the solution by plane $w=0$ are shown. The solutions correspond to 
      $s_{u}=s_{v}=s_{w}=3$ and 7 cells in each velocity direction. 
    (a) $t=0$ ns, (b) $t=0.5$ ns, (c) $t=2.1$ ns, and (d) $t=4.5$ ns. }
\end{figure}

\begin{figure}[tp]
  \begin{tabular}{@{}cc@{}}
  \includegraphics[height=.24\textheight]{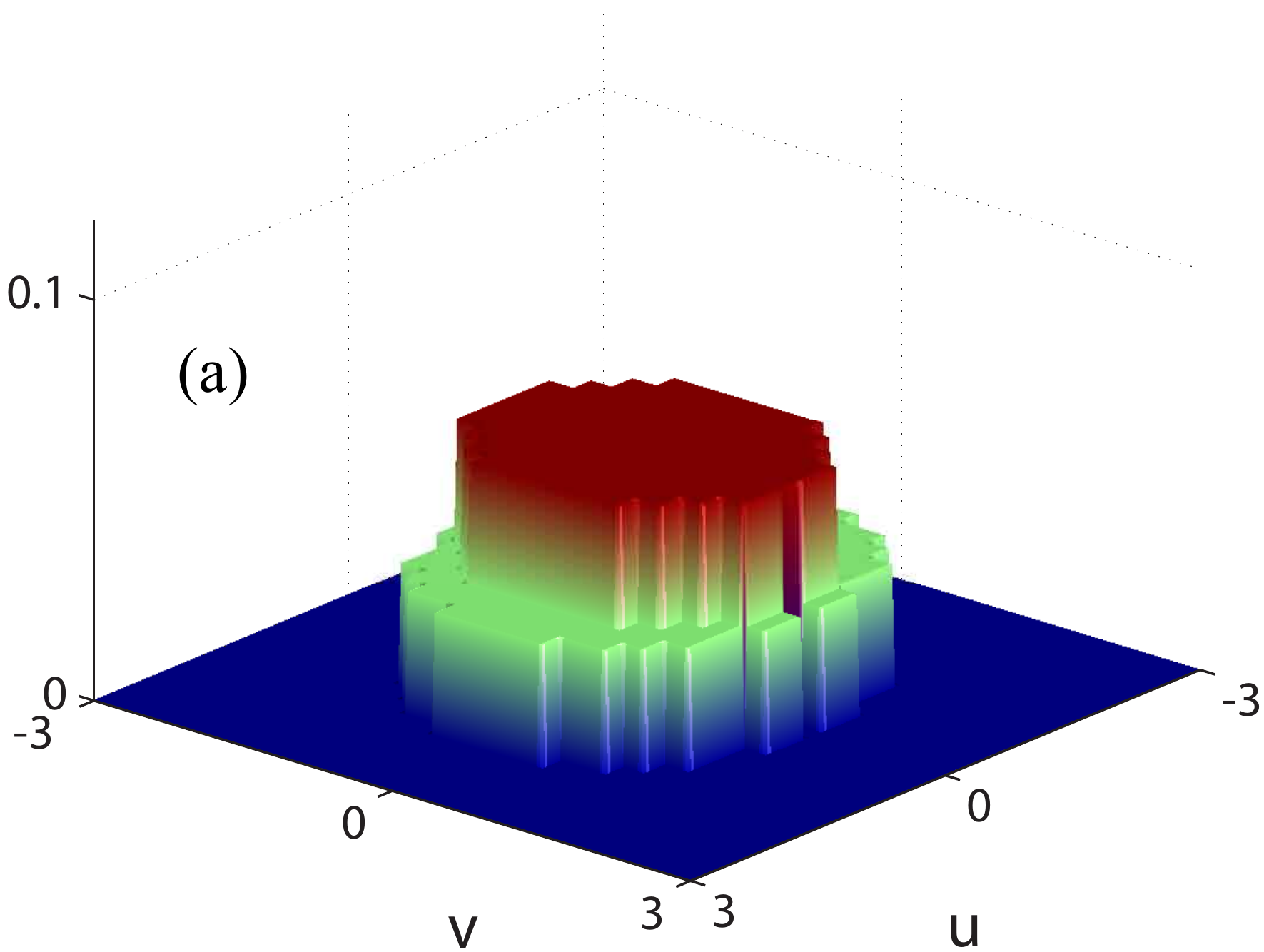}&
  \includegraphics[height=.24\textheight]{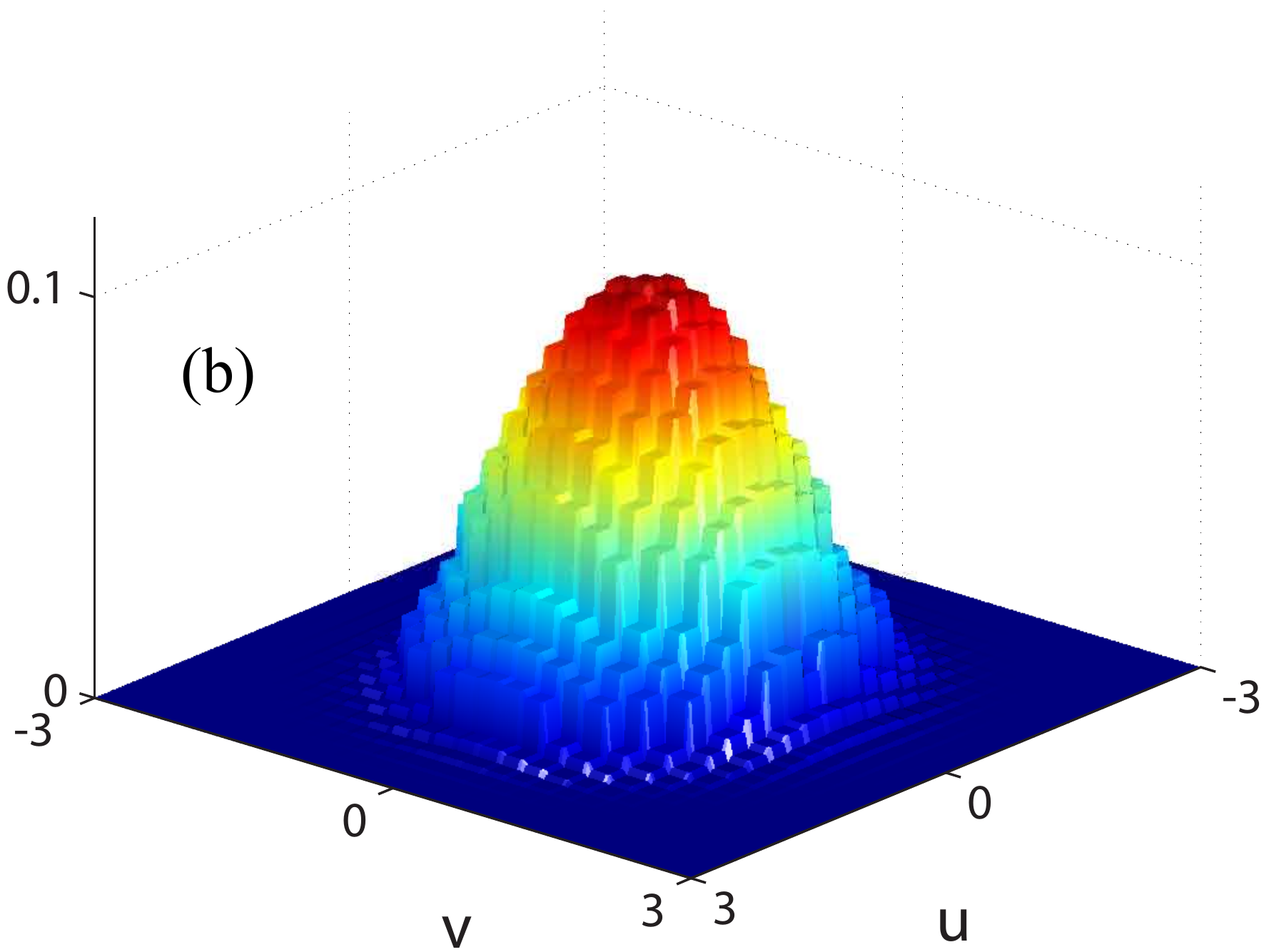}\\
  \includegraphics[height=.24\textheight]{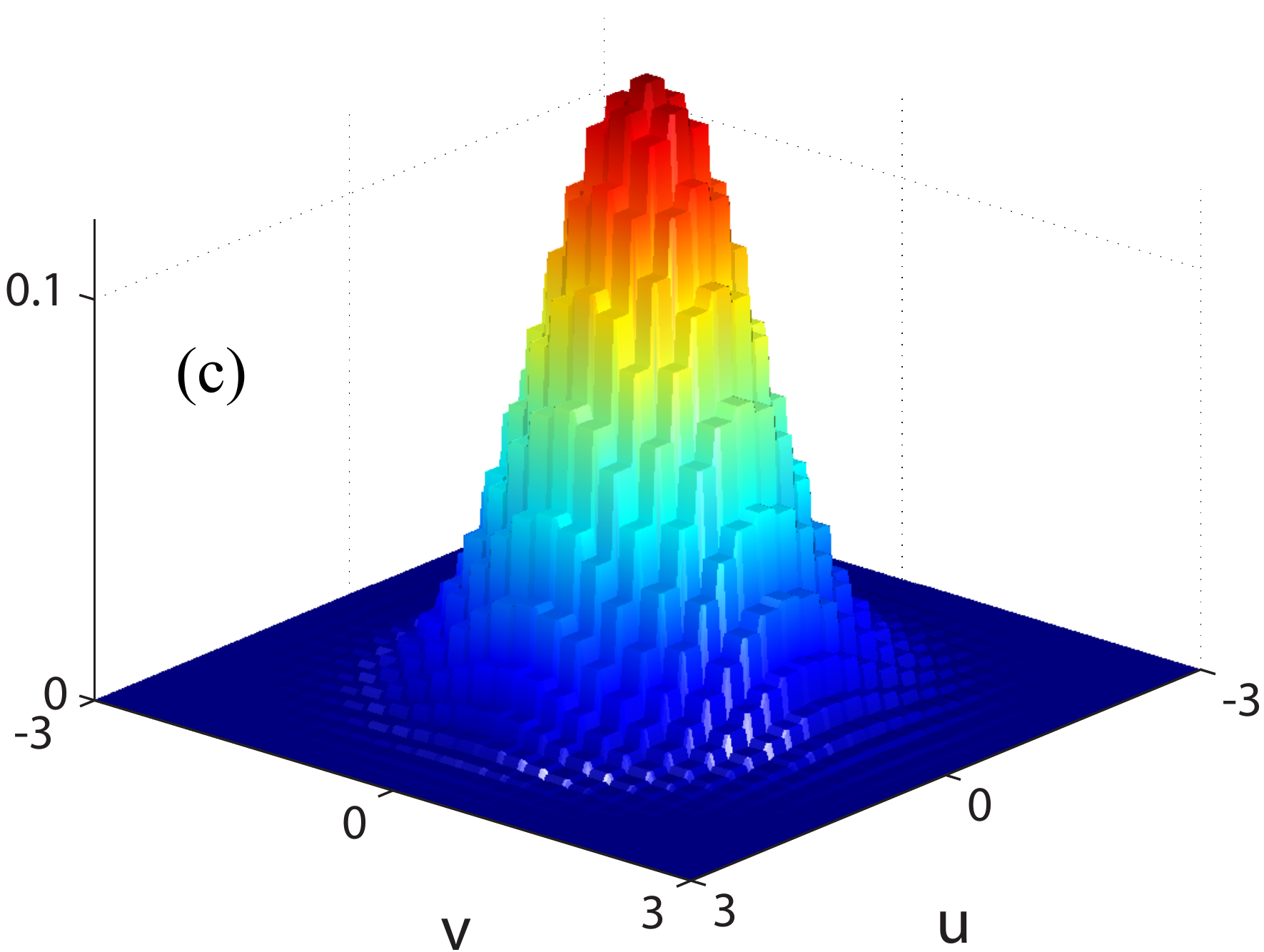}&
  \includegraphics[height=.24\textheight]{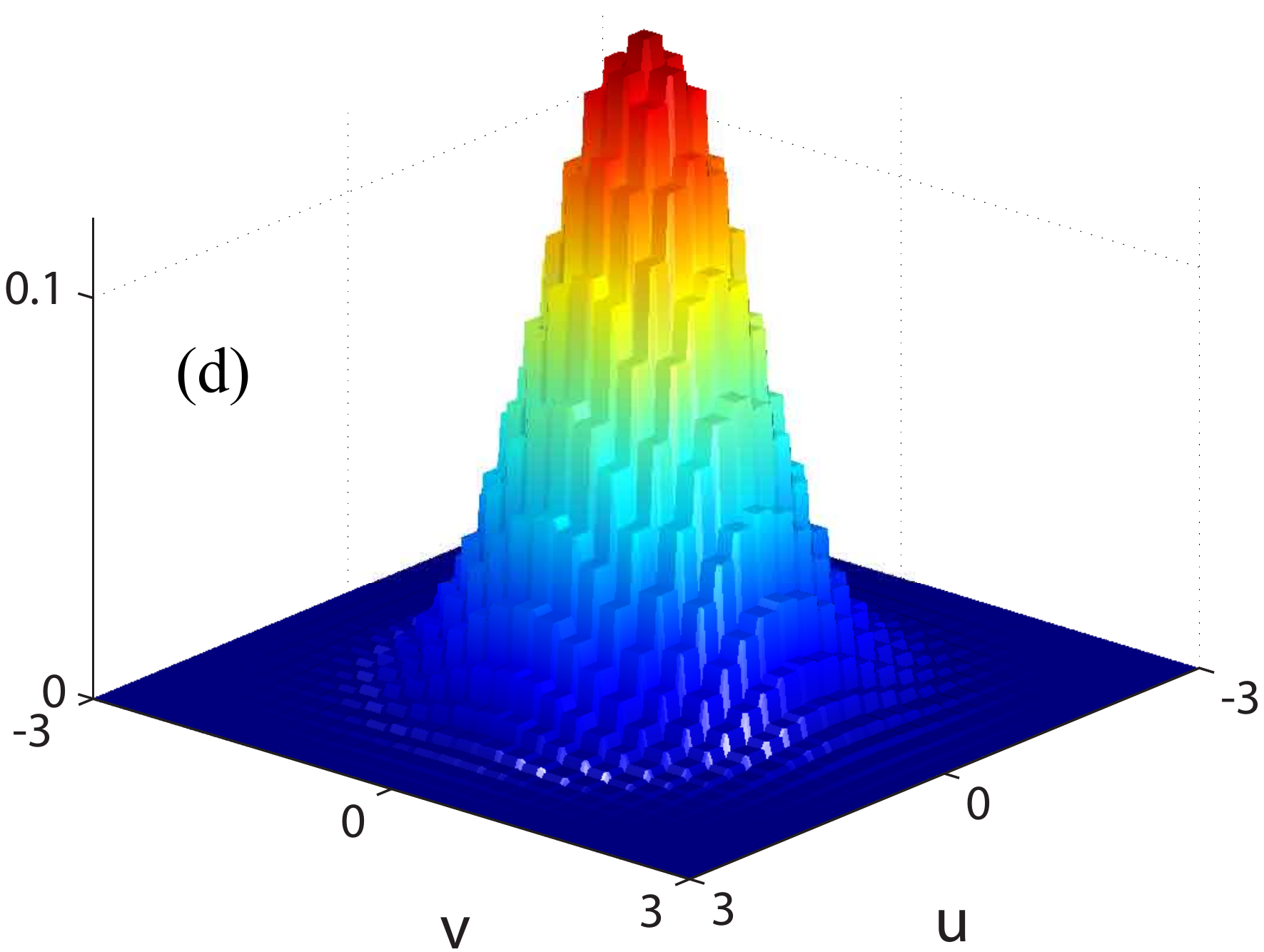}\\
  \end{tabular}
  \caption{\label{fig04a} Relaxation of two artificial streams using piece-wise constant DG approximations. 
    Sections of the solution by plane $w=0$ are shown. The solutions correspond to  
    $s_{u}=s_{v}=s_{w}=1$ and 27 cells in each velocity direction. (a) $t=0$ ns; (b) $t=0.5$ ns;
    (c) $t=2.1$ ns; and (d) $t=4.5$ ns.}
\end{figure}

Our second problem is concerned with the relaxation of two artificial streams with discontinuous initial 
data. Specifically, the initial distribution is a sum of two functions of the form 
\begin{equation*}
f_{\mathrm{X}}(\vec{v})= 
\left\{ \begin{array}{lr}
\rho h, & \mbox{if} \ |\vec{v}-\vec{\bar{v}}| \leq r\, , \\
0, & \mbox{if} \ |\vec{v}-\vec{\bar{v}}| > r\, ,
\end{array} \right. \ , \quad r=(T/5R)^{1/2}, \ \mbox{and}\ h=(15\sqrt{5}/4\pi) (R/T)^{3/2}. 
\end{equation*} 
Here $\rho$, $\vec{\bar{v}}$ and $T$ are given parameters. It is a straightforward exercise 
to verify that $\rho$, $\vec{\bar{v}}$ and $T$ coincide with the mass density, bulk velocity and temperature of the distribution 
$f_{\mathrm{X}}(\vec{v})$, correspondingly.  
The gas is argon. The values of the macroparameters for the two artificial streams are $\rho_{1}=0.332$ kg/m${}^3$, 
$\vec{\bar{v}}_{1}= (106.0, 0, 0)$ m/s, $T_{1}=300$ K and $\rho_{2}=0.332$ kg/m${}^3$, 
$\vec{\bar{v}}_{2}= (-106.0, 0, 0)$ m/s, $T_{2}=300$ K. Graphs of two-dimensional cross-sections of the initial 
distribution by plane $w=0$ are given in Figures~\ref{fig04}(a) and \ref{fig04a}(a). The mean time between molecular 
collisions is estimated to be about $0.39$~ns. The simulations are performed for $190$ ns. In Figure~\ref{fig04} the simulations 
are presented using piece-wise quadratic approximations, $s_{u}=s_{v}=s_{w}=3$, on $M=7$ cells in each velocity dimension and in Figure~\ref{fig04a} 
using piece-wise constant approximations, $s_{u}=s_{v}=s_{w}=1$, on $M=27$ cells in each velocity dimension. One can see that the simulations 
are consistent in capturing the process of relaxation. The approximation artifacts visible in Figure~\ref{fig04}(a) are caused 
by insufficient resolution. These artifacts subside as the distribution functions relaxes and gets smoother. 

\begin{figure}[tp]
  \begin{tabular}{@{}c@{}c@{}}
  \includegraphics[height=.29\textheight]{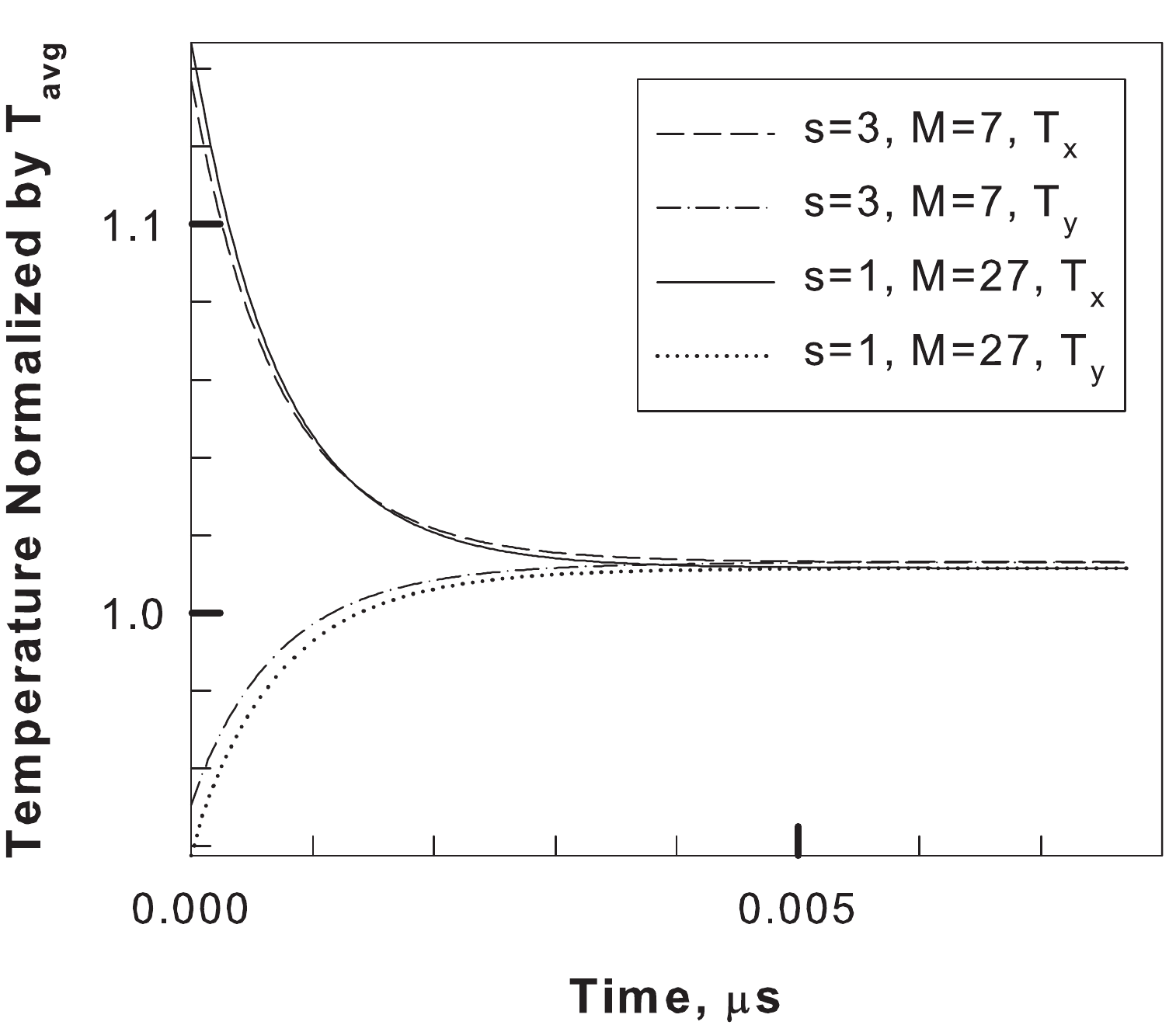} &
  \includegraphics[height=.29\textheight]{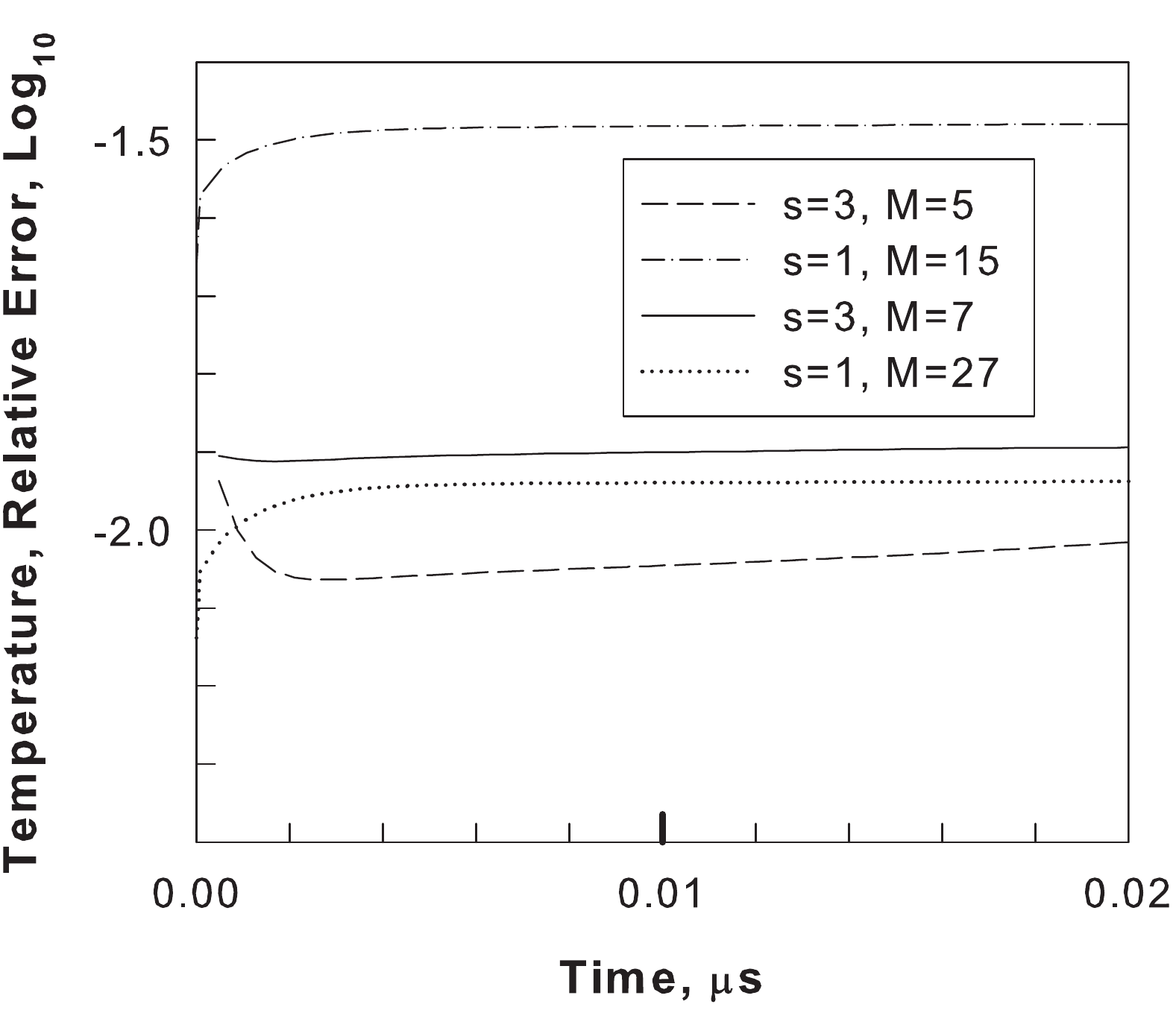} \\
  {\small (a) }&{\small (b) } 
  \end{tabular}
  \caption{\label{fig05} Relaxation of two artificial streams. (a) Ratios $T_{x}/T_{\mathrm{avg}}$ and 
  $T_{y}/T_{\mathrm{avg}}$ are shown. Here $T_{x}$ and $T_{y}$ are directional temperatures and 
  $T_{\mathrm{avg}}=T/3$. (b) Relative error in the solutions temperature. Piece-wise constant and piece-wise quadratic 
  DG approximations are compared. Fast growth of error occurs in the piece-wise constant approximation at 
  the initial stages of relaxation.}
\end{figure}

In Figure~\ref{fig05}(b) the conservation of temperature in the DG solutions is given in the cases of $s_{u}=s_{v}=s_{w}=1$ and 
$s_{u}=s_{v}=s_{w}=3$. One can notice that the approximations of discontinuous initial data are less accurate and 
also that the piece-wise constant DG approximations perform inferior to the piece-wise quadratic approximations. 
This observation is consistent with studies performed in \cite{Alekseenko2011}. 
It is believed that the fast convergence of the piece-wise constant approximations in the problem of relaxation of 
two Maxwellian distributions was due to the solution smoothness and to the fact that kinetic solutions exponentially 
decrease at infinity. As can be seen from the second example, accuracy of integration in piece-wise constant approximations 
deteriorates dramatically if the approximated solution is not smooth. One can see that growth of numerical errors in 
piece-wise constant DG approximations is fastest at the initial stages of the relaxation, when the solution is still far 
from being smooth. As the solution is getting smoother, however, the accuracy of quadrature rules increases and the 
conservation laws are satisfied with a better accuracy. As a result, we observe almost no change in 
the errors once the solutions reaches the steady state.

\section{Conclusion}

We developed a discontinuous Galerkin discretization of the Boltzmann equation in the velocity space 
using a symmetric bilinear form of the Galerkin projection of the collision operator. The time-independent kernel 
of the bilinear operator carries the information about the geometry of the velocity discretization and about the 
collision model. Properties of the kernel were studies and several practical statements about the kernel symmetry were 
formulated. Evaluation of the kernel was implemented using an MPI parallelization algorithm that scales to a 
large number of processors. The discontinuous Galerkin approach was applied to the problem of spatially homogeneous 
relaxation. Discretizations with up to $M=33$ degrees of freedom per velocity dimension were successfully tested on a 
single processor. DG solutions to the problem of spatially homogeneous relaxation showed excellent agreement with 
the solutions obtained by established DSMC codes. The solutions conserve mass momentum and temperature with 
a good accuracy.

The main obstacles to increasing the velocity resolution to hundreds of nodes per velocity dimension 
are the large amount of components per velocity basis function in the pre-computed collision kernel 
and the large number of resulting arithmetic operations to evaluate the collision integral. 
These numbers were demonstrated to grow as $O(n^5)$ and $O(n^8)$, respectively for a single spatial cell. 
To overcome this problem, the authors are developing algorithms for parallelization 
of the evaluation of the collision integral to up to thousands of processors. An additional improvement is 
expected by constructing efficient Galerkin basis so as to minimise the total number of degrees of freedom in velocity 
variable.  The discrete Galerkin form of the Boltzmann equation 
offers unique insights to the numerical properties of its solution. In the future, the authors will explore the 
eigenstructure of the bilinear discrete collision kernel and connect the new knowledge to the familiar numerical 
properties of the solutions to the Boltzmann equation. It is anticipated that this study will lead to the construction 
of more efficient approximation techniques.

\section*{Acknowledgement}

The first author was supported by NRC Resident Associate Program in 2011-2012. The authors would like to 
thank Professor I.\ Boyd and Dr.\ J.\ Burt for their interest in this work and for inspiring discussions. 
The authors also thank Professors S.\ Gimelshein, J.\ Shen, B.\ Lucier and V.\ Panferov for their interest 
in this work. The computer resources were provided by the U.S.\ Department of Defense, High
Performance Computing, Defence Shared Resource Center at AFRL, Wright-Patterson AFB, Ohio. 
Additional computational resources were provided by the Rosen Center for Advanced Computing at Purdue University
and by the Department of Mathematics at Purdue University. 
The authors cordially thank Dr.\ L.L.\ Foster for her help in proofreading the paper.



\appendix
\section{Proofs of Theorems~\ref{thm2.1}--\ref{thm2.3}} 
\label{AppendixA}

\begin{lem21} 
Let operator $A(\vec{v},\vec{v}_{1};\phi^{j}_{i})$ be defined by (\ref{eq2.4}) with all gas particles having
the same mass and the potential of the particles interaction being spherically symmetric. 
Then $A(\vec{v},\vec{v}_{1};\phi^{j}_{i})$ is symmetric with respect to $\vec{v}$ and $\vec{v}_{1}$, that is 
\begin{equation*}
A(\vec{v},\vec{v}_{1};\phi^{j}_{i}) = A(\vec{v}_{1},\vec{v};\phi^{j}_{i}), \qquad \forall \vec{v},\vec{v}_{1}\in \R^3.
\end{equation*} 
Also,
\begin{equation*}
A(\vec{v},\vec{v};\phi^{j}_{i}) = 0, \qquad \forall \vec{v} \in \R^3.
\end{equation*} 
\end{lem21}

\proof
Formula (\ref{eq2.6}) follows immediately from (\ref{eq2.4}) by noticing that if $\vec{v}=\vec{v}_{1}$ then $\vec{g}=0$. Therefore, (\ref{eq2.4}) is automatically zero.

To prove (\ref{eq2.5}) we recall that in the case when all particles have the same mass, 
the interchange of pre-collision velocities $\vec{v}$ and $\vec{v}_{1}$ will result in 
the interchange of post collision velocities $\vec{v}'$ and $\vec{v}'_{1}$. In particular, 
no different post-collision velocities result from the interchange of $\vec{v}$ and 
$\vec{v}_{1}$. The statement follows by noticing the 
symmetry of the expression under the integral (\ref{eq2.5}) is both pairs of velocities.

For a more ``mathematical'' proof one may recall that in 
the case of a spherically symmetric potential the pre-collision velocities $\vec{v}$ and 
$\vec{v}_{1}$ and the post-collision velocities $\vec{v}'$ and $\vec{v}'_{}$ of the molecules 
are located on a sphere with the center at $(\vec{v}+\vec{v}_{1})/2$ and the radius of 
$|\vec{g}|/2$, see for example \cite{Cercignani2000}. Furthermore, from some geometrical and 
physical considerations on concludes that $\vec{v}$, $\vec{v}_{1}$, $\vec{v}'$ and 
$\vec{v}'_{}$ form a plain rectangle. Then the statement follows by observing that 
the rectangle is determined by $\vec{v}$ and $\vec{v}_{1}$ in a unique way.)
\endproof

\begin{lem22} 
Let operator $A(\vec{v},\vec{v}_{1};\phi^{j}_{i})$ be defined by (\ref{eq2.4}) and let 
the potential of molecular interaction be dependent only on the distance between the particles. Then $\forall\xi\in \R^3$
\begin{equation*}
A(\vec{v}+\vec{\xi},\vec{v}_{1}+\vec{\xi};\phi^{j}_{i}(\vec{v}-\vec{\xi}))=
A(\vec{v},\vec{v}_{1};\phi^{j}_{i}) \, .
\end{equation*}
\end{lem22}

\proof
Consider $A(\vec{v}+\vec{\xi},\vec{v}_{1}+\vec{\xi};\phi^{j}_{i}(\vec{v}-\vec{\xi}))$. We clarify that these notations mean that
vectors $\vec{v}$ and $\vec{v}_{1}$ in (\ref{eq2.4}) are replaced with $\vec{v}+\vec{\xi}$ and $\vec{v}_{1}+\vec{\xi}$
correspondingly and that basis function $\phi^{j}_{i}(\vec{v})$ is replaced with a ``shifted'' function 
$\phi^{j}_{i}(\vec{v}-\vec{\xi})$. We notice that the relative speed of the molecules with velocities $\vec{v}+\vec{\xi}$ 
and $\vec{v}_{1}+\vec{\xi}$ is still $\vec{g}=\vec{v}+\vec{\xi}-(\vec{v}_{1}+\vec{\xi}_{1})=\vec{v}-\vec{v}_{1}$. 
Since the potential of molecular interaction depends only on the distance between the particles and in particular 
it does not depend on the particles individual velocities, the post-collision velocities will be $\vec{v}'+\vec{\xi}$ and 
$\vec{v}'_{1}+\vec{\xi}$. The rest of the statement follows a direct substitution:
\begin{align*}
A(\vec{v}&+\vec{\xi},\vec{v}_{1}+\vec{\xi};\phi^{j}_{i}(\vec{v}-\vec{\xi})) \\
& = \frac{|\vec{g}|}{2} \int_{0}^{2\pi} \int_{0}^{b_{\ast}}
(\phi^{j}_{i}((\vec{v}'+\vec{\xi})-\vec{\xi})+\phi^{j}_{i}((\vec{v}'_{1}+\vec{\xi})-\vec{\xi})
-\phi^{j}_{i}((\vec{v}+\vec{\xi})-\vec{\xi})-\phi^{j}_{i}((\vec{v}_{1}+\vec{\xi})-\vec{\xi}) b\, db\, d\varepsilon \, \\
& = \frac{|\vec{g}|}{2} \int_{0}^{2\pi} \int_{0}^{b_{\ast}} 
(\phi^{j}_{i}(\vec{v}')+\phi^{j}_{i}(\vec{v}'_{1})-\phi^{j}_{i}(\vec{v})-\phi^{j}_{i}(\vec{v}_{1})) b\, db\, d\varepsilon \,\\
& = A(\vec{v},\vec{v}_{1};\phi^{j}_{i})\, . 
\end{align*}
\endproof

\begin{lem23} 
Let operator $A(\vec{v},\vec{v}_{1};\phi^{j}_{i})$ be defined by (\ref{eq2.4}) and let 
the potential of molecular interaction be dependent only on the distance between the particles. Let $S:\R^3\to \R^{3}$ 
be a linear isometry of $\R^{3}$. Then 
\begin{equation*}
A(S\vec{v},S\vec{v}_{1};\phi^{j}_{i}(S^{-1}\vec{u})=
A(\vec{v},\vec{v}_{1};\phi^{j}_{i}) \, .
\end{equation*}
\end{lem23}

\proof
We begin with a remark that will be helpful to highlight the 
mathematical structure of the proof. In kinetic theory there is a 
connection between the velocity space and the physical space in that 
velocity vectors are describing motion in the physical space and 
that displacement vectors in physical space are naturally identified 
with points in the velocity space. By identifying this 
connection, we implicitly introduce an angle and orientation preserving 
affine transformation that maps the first axis of the positive 
coordinate triple in the physical space to the first axis of the positive 
coordinate triple in the velocity space, the second axis into the second one 
and the third axis into the third one. As a result, we obtain a way to map 
vectors from the physical space to the velocity space and vice-versa. Connection 
between physical and velocity spaces can also be seen in definition 
(\ref{eq2.4}). Indeed, post-collision velocities 
$\vec{v}'$ and $\vec{v}'_{1}$ are functions of the pre-collision velocities, $\vec{v}$ and $\vec{v}_{1}$, 
the impact parameters $b$ and $\varepsilon$ and the collision model. While the former belongs to 
the velocity space, the latter belongs to the physical space. Moreover, to define parameters $b$ of 
the molecular shortest approach distance and $\varepsilon$ of the relative angle of the collision plane,
we must move the vector of relative velocity $\vec{g}=\vec{v}-\vec{v}_{1}$ into the physical space, 
e.g., \cite{Kogan1969}, Section 1.3. Specifically, we define $b$ as the 
distance from the center of the second molecule to the line coming through the center of the first 
molecule in the direction of their relative velocity $\vec{g}$. Also, the parameter 
$\varepsilon$ is the angle between the collision plane formed by the image of the 
vector $\vec{g}$ in the physical space and the centers of the 
colliding molecules and a reference coordinate plane. Thus, 
to evaluate (\ref{eq2.4}) we need to map vectors between physical and 
velocity spaces many times. Applying this reasoning to the statement of 
the theorem in question, we notice that the left side of (\ref{eq2.8}) 
is evaluated relative to velocities $S\vec{v}$ and $S\vec{v}_{1}$ 
while the right side relative to velocities $\vec{v}$ and $\vec{v}_{1}$. 
Therefore separate sets of impact parameters are introduced for each case. 
What we are intended to show is that the result is the same for both cases.
 
Applying definition (\ref{eq2.4}) to the left side of (\ref{eq2.8}) we have 
\begin{align}
\label{eq2.9}
A(S\vec{v},&S\vec{v}_{1};\phi^{j}_{i}(S^{-1}\vec{u})) \nonumber \\
&= \frac{|\vec{\tilde{g}}|}{2} \int_{0}^{2\pi} \int_{0}^{b_{\ast}}
[\phi^{j}_{i}(S^{-1}(S\vec{v})'))+\phi^{j}_{i}(S^{-1}(S\vec{v}_{1})')
-\phi^{j}_{i}(S^{-1}S\vec{v})-\phi^{j}_{i}(S^{-1}S\vec{v}_{1})] \tilde{b}\, d\tilde{b}\, d\tilde{\varepsilon} \, \nonumber \\
&=\frac{|\vec{g}|}{2} \int_{0}^{2\pi} \int_{0}^{b_{\ast}}
[\phi^{j}_{i}(S^{-1}(S\vec{v})')+\phi^{j}_{i}(S^{-1}(S\vec{v}_{1})')
-\phi^{j}_{i}(\vec{v})-\phi^{j}_{i}(\vec{v}_{1})] \tilde{b}\, d\tilde{b}\, d\tilde{\varepsilon} \, ,
\end{align} 
where $\vec{\tilde{g}}=S\vec{v}-S\vec{v}_{1}$ and  $\tilde{b}$ and $\tilde{\varepsilon}$ are 
the impact parameters defined in the local coordinate system of the molecules with  
velocities $S\vec{v}$ and $S\vec{v}_1$. Notice that because $S$ is linear, we have
$\vec{\tilde{g}}=S\vec{v}-S\vec{v}_{1}=S(\vec{v}-\vec{v}_{1})=S\vec{g}$. Also, 
$S$ is an isometry, therefore, $|\vec{\tilde{g}}|=|S \vec{g}|=|\vec{g}|$. 

Let us now describe the local coordinate systems that are introduced for both pairs of molecules: 
the pair with velocities $S\vec{v}$ and $S\vec{v}_{1}$ and the pair with velocities $\vec{v}$ 
and $\vec{v}_{1}$. Let us
consider the molecules with velocities $S\vec{v}$ and $S\vec{v}_{1}$, first. According to the formalism
described above, we define $\tilde{b}$ as the distance from molecule with 
velocity $S\vec{v}_{1}$ to the 
line passing through the center of molecule with velocity $S\vec{v}$ in the direction of $\vec{\tilde{g}}$.
We let the coordinate system be introduced for the pair of colliding molecules with velocities $S\vec{v}$
and $S\vec{v}_{1}$ so that its origin located at the center of the molecule with velocity $S\vec{v}$ and its 
$\tilde{\xi}_{1}$ axis directed along $\vec{\tilde{g}}$. Then the parameter 
$\tilde{b}$ is the distance to the $\tilde{\xi}_{1}$ axis. The parameter $\tilde{\varepsilon}$ 
of the angle of the collision plane is defined relevant to a reference plane. We let 
axes $\tilde{\xi}_{2}$ and $\tilde{\xi}_{3}$ be selected to make a right triple and designate 
the plane $\tilde{\xi}_{1}\tilde{\xi}_{2}$ as the reference plane. We note that the collision plane 
contains the $\tilde{\xi}_{1}$ axis and let the angle $\tilde{\varepsilon}$ between the reference 
plane and the collision plane be measured from $\tilde{\xi}_{2}$ axis toward the $\tilde{\xi}_{3}$ 
axis. Notice that these assumptions do not limit the generality of the argument because any 
admissible parametrization should produce the correct value of the integral above. 

We will now show that a choice of the parameter $\varepsilon$ can be made 
in a local system of coordinates corresponding to the pair of molecules colliding with 
velocities $\vec{v}$ and $\vec{v}_{1}$, such that $\vec{v}' = S^{-1}(S\vec{v})'$ 
and $\vec{v}'_1=S^{-1}(S\vec{v}_1)'$ as long as $b=\tilde{b}$ and 
$\varepsilon=\tilde{\varepsilon}$. We let the origin of the second set of coordinates 
be located at the center of molecule with velocity $\vec{v}$ and its ${\xi}_{1}$ axis 
directed along ${\vec{g}}$. Then parameter $b$ gives the distance from the molecule with velocity 
$\vec{v}_{1}$ to the axis $\xi_{1}$. We define axes ${\xi}_{2}$ 
and ${\xi}_{3}$ to be the images of $\tilde{\xi}_{2}$ and $\tilde{\xi}_{3}$, respectively, 
under the action of $S^{-1}$. Specifically, using the mapping between the physical space and 
the velocity space we consider the unit vectors giving the directions of the axes $\tilde{\xi}_{2}$ 
and $\tilde{\xi}_{3}$ in the velocity space. Let these vectors be $\vec{\tilde{\tau}}_{2}$ 
and $\vec{\tilde{\tau}}_{3}$, respectively. Applying the inverse transformation $S^{-1}$ to these vectors 
and moving back to physical space, we require that $\vec{\tau}_{2}=S^{-1}\vec{\tilde{\tau}}_{2}$ and
$\vec{\tau}_{3}=S^{-1}\vec{\tilde{\tau}}_{3}$ be the unit vectors of the axes $\xi_{2}$ and $\xi_{3}$.
It is a simple check that $\vec{\tau}_{1}=S^{-1}\vec{\tilde{\tau}}_{1}$, where 
$\vec{\tilde{\tau}}_{1}=\vec{\tilde{g}}/|\vec{\tilde{g}}|$ defines the $\tilde{\xi}_{1}$ axis.

Since the molecular potential is spherically symmetric, the relative velocity undergoes 
a specular reflection on the impact, see e.g. \cite{Cercignani2000}, Section 1.2. Furthermore, 
the trajectories of the colliding molecules are contained in the collision plane, see 
e.g., \cite{Kogan1969}, Section 1.3. Thus the post-collision relative velocity belongs to the
collision plane. Let us denote by $O[b,\varepsilon,|\vec{g}|](\vec{g}):\R^{3}\to \R^{3}$ 
the operator of flat rotation of the vector of relative velocity in collision plane. 
Notice that the rotation angle depends on the approach distance $b$, the collision model 
and on the norm of the relative velocity. It, however, does not depend on the direction 
of $\vec{g}$. Re-writing slightly the familiar formulas for the post-collision 
velocities (see, e.g., \cite{Cercignani2000}, Section 1.2), we have
\begin{align*}
\vec{v}'&=\vec{v}+(O[b,\varepsilon,|\vec{g}|](\vec{g})-\vec{g})/2\, ,\\
\vec{v}'_{1}&=\vec{v}_{1} - (O[b,\varepsilon,|\vec{g}|](\vec{g})-\vec{g})/2\, .\\
\end{align*}
Similarly the molecules with velocities $S\vec{v}$ and $S\vec{v}_{1}$ we have 
\begin{align*}
(S\vec{v})'& = S\vec{v}
+ (O[\tilde{b},\tilde{\varepsilon},|\vec{\tilde{g}}|](\vec{\tilde{g}})-\vec{\tilde{g}})/2\, ,\\
(S\vec{v}_{1})'& = S\vec{v}_{1} 
- (O[\tilde{b},\tilde{\varepsilon},|\vec{\tilde{g}}|](\vec{\tilde{g}})-\vec{\tilde{g}})/2\, .\\
\end{align*}
recalling that $|\vec{\tilde{g}}|=|\vec{g}|$, we obtain
\begin{align}
\label{eq2.10}
(S\vec{v})'& = S\vec{v}
+ (O[\tilde{b},\tilde{\varepsilon},|\vec{{g}}|](\vec{\tilde{g}})-\vec{\tilde{g}})/2\, ,\nonumber \\
(S\vec{v}_{1})'& = S \vec{v}_{1} 
- (O[\tilde{b},\tilde{\varepsilon},|\vec{{g}}|](\vec{\tilde{g}})-\vec{\tilde{g}})/2\, .
\end{align}
We now consider the two vectors $O[\tilde{b},\tilde{\varepsilon},|\vec{{g}}|](\vec{\tilde{g}})$
and $S(O[b,\varepsilon,|\vec{g}|](\vec{g}))$. Because $S$ is a linear isometry, it preserves 
angles between vectors. Therefore, for any $0<n<b^{\ast}$ the plane formed by the vectors
$S(O[b,\varepsilon,|\vec{g}|](\vec{g}))$, $\vec{\tilde{g}}=S\vec{g}$ will be making 
angle $\varepsilon$ with the plane formed by vectors $\vec{\tilde{\tau}}_{2}=S\vec{{\tau}}_{2}$ and 
$\vec{\tilde{\tau}}_{3}=S\vec{{\tau}}_{3}$. Moreover, $S(O[b,\varepsilon,|\vec{g}|](\vec{g}))$ 
will make the same angle with vector
$\vec{\tilde{g}}=S\vec{g}$ as $O[b,\varepsilon,|\vec{g}|](\vec{g})$ is making with $\vec{g}$. 
Recalling that the angle of flat rotation in (\ref{eq2.10}) only depends on $\tilde{b}$ we conclude that
if $\tilde{b}=b$ then $O[\tilde{b},\tilde{\varepsilon},|\vec{{g}}|](\vec{\tilde{g}})$
will make the same angle with $\vec{\tilde{g}}$ as $O[b,\varepsilon,|\vec{g}|](\vec{g})$ is making 
with $\vec{g}$. Noting that as long as $\tilde{\varepsilon}=\varepsilon$, both 
$S(O[b,\varepsilon,|\vec{g}|](\vec{g}))$ and $O[\tilde{b},\tilde{\varepsilon},|\vec{{g}}|](\vec{\tilde{g}})$
lay in the same plane, we conclude that they are the same vectors. Therefore, 
$S(O[b,\varepsilon,|\vec{g}|](\vec{g})) = O[\tilde{b},\tilde{\varepsilon},|\vec{{g}}|](\vec{\tilde{g}})$
and $O[b,\varepsilon,|\vec{g}|](\vec{g}) =
S^{-1}(O[\tilde{b},\tilde{\varepsilon},|\vec{{g}}|](\vec{\tilde{g}}))$ as long as 
$\tilde{b}=b$ and $\tilde{\varepsilon}=\varepsilon$. By applying $S^{-1}$ to both sides of (\ref{eq2.8})
we conclude that $\vec{v}' = S^{-1}(S\vec{v})'$ 
and $\vec{v}'_1=S^{-1}(S\vec{v}_1)'$ as long as $b=\tilde{b}$ and 
$\varepsilon=\tilde{\varepsilon}$. The rest of the statement follows by introducing the substitution 
$\tilde{b}=b$ and $\tilde{\varepsilon}=\varepsilon$ in (\ref{eq2.9}) and replacing  $S^{-1}(S\vec{v})'$ 
and $S^{-1}(S\vec{v}_1)'$ with  $\vec{v}'$ and $\vec{v}'_1$, respectively.
\endproof

We notice that in the case of a general molecular interaction potential, the angle of 
rotation in the collision plane included in $O[\tilde{b},\tilde{\varepsilon},|\vec{{g}}|]$,
depends on the relative velocity. Because of this, it is unlikely that Lemma~\ref{thm2.3} can be
extended to a contraction mapping $S$. We however anticipate that it is 
possible to develop analogs of formulas (\ref{eq2.8}) by introducing an appropriate scaling 
parameter in the case of molecular potentials that do not depend on the relative velocity, 
e.g., for hard spheres. Such formulas would alleviate evaluation and storage 
of $A(\vec{v},\vec{v}_{1};\phi^{j}_{i})$.



\bibliographystyle{model1-num-names}
\bibliography{n11122012}

\end{document}